\documentclass[twoside,twocolumn,9pt]{article}
\usepackage{extsizes}
\usepackage[super,sort&compress,comma]{natbib} 
\usepackage[version=3]{mhchem}
\usepackage[left=1.5cm, right=1.5cm, top=1.785cm, bottom=2.0cm]{geometry}
\usepackage{balance}
\usepackage{mathptmx}
\usepackage{sectsty}
\usepackage{graphicx} 
\usepackage{tabularx}
\usepackage{rotating}
\usepackage{lastpage}
\usepackage[format=plain,justification=justified,singlelinecheck=false,font={stretch=1.125,small,sf},labelfont=bf,labelsep=space]{caption}
\usepackage{float}
\usepackage{fancyhdr}
\usepackage{fnpos}
\usepackage[english]{babel}
\addto{\captionsenglish}{%
  
}
\usepackage{array}
\usepackage[normalem]{ulem}
\usepackage{droidsans}
\usepackage{charter}
\usepackage[T1]{fontenc}
\usepackage[usenames,dvipsnames]{xcolor}
\usepackage{setspace}
\usepackage[compact]{titlesec}
\usepackage{hyperref}

\def\bk{{\bf k}}
\def\bq{{\bf q}}
\usepackage{amsmath}
\usepackage{amssymb}

\usepackage{epstopdf}

\definecolor{cream}{RGB}{222,217,201}

\begin{document}

\pagestyle{fancy}
\thispagestyle{plain}
\fancypagestyle{plain}{
\renewcommand{\headrulewidth}{0pt}
}

\makeFNbottom
\makeatletter
\renewcommand\LARGE{\@setfontsize\LARGE{15pt}{17}}
\renewcommand\Large{\@setfontsize\Large{12pt}{14}}
\renewcommand\large{\@setfontsize\large{10pt}{12}}
\renewcommand\footnotesize{\@setfontsize\footnotesize{7pt}{10}}
\makeatother

\renewcommand{\thefootnote}{\fnsymbol{footnote}}
\renewcommand\footnoterule{\vspace*{1pt}%
\color{cream}\hrule width 3.5in height 0.4pt \color{black}\vspace*{5pt}} 
\setcounter{secnumdepth}{5}

\makeatletter 
\renewcommand\@biblabel[1]{#1}            
\renewcommand\@makefntext[1]%
{\noindent\makebox[0pt][r]{\@thefnmark\,}#1}
\makeatother 
\renewcommand{\figurename}{\small{Fig.}~}
\sectionfont{\sffamily\Large}
\subsectionfont{\normalsize}
\subsubsectionfont{\bf}
\setstretch{1.125} 
\setlength{\skip\footins}{0.8cm}
\setlength{\footnotesep}{0.25cm}
\setlength{\jot}{10pt}
\titlespacing*{\section}{0pt}{4pt}{4pt}
\titlespacing*{\subsection}{0pt}{15pt}{1pt}

\fancyfoot{}
\fancyfoot[LO,RE]{\vspace{-7.1pt}\includegraphics[height=9pt]{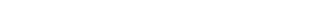}}
\fancyfoot[CO]{\vspace{-7.1pt}\hspace{13.2cm}\includegraphics{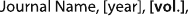}}
\fancyfoot[CE]{\vspace{-7.2pt}\hspace{-14.2cm}\includegraphics{head_foot/RF}}
\fancyfoot[RO]{\footnotesize{\sffamily{1--\pageref{LastPage} ~\textbar  \hspace{2pt}\thepage}}}
\fancyfoot[LE]{\footnotesize{\sffamily{\thepage~\textbar\hspace{3.45cm} 1--\pageref{LastPage}}}}
\fancyhead{}
\renewcommand{\headrulewidth}{0pt} 
\renewcommand{\footrulewidth}{0pt}
\setlength{\arrayrulewidth}{1pt}
\setlength{\columnsep}{6.5mm}
\setlength\bibsep{1pt}

\makeatletter 
\newlength{\figrulesep} 
\setlength{\figrulesep}{0.5\textfloatsep} 

\newcommand{\topfigrule}{\vspace*{-1pt}%
\noindent{\color{cream}\rule[-\figrulesep]{\columnwidth}{1.5pt}} }

\newcommand{\botfigrule}{\vspace*{-2pt}%
\noindent{\color{cream}\rule[\figrulesep]{\columnwidth}{1.5pt}} }

\newcommand{\dblfigrule}{\vspace*{-1pt}%
\noindent{\color{cream}\rule[-\figrulesep]{\textwidth}{1.5pt}} }

\makeatother

\newcommand{\CT}[1]{\textcolor{blue}{#1}}
\newcommand{\stc}[1]{\textcolor{black}{\sout{#1}}} 

\twocolumn[
  \begin{@twocolumnfalse}
{\includegraphics[height=30pt]{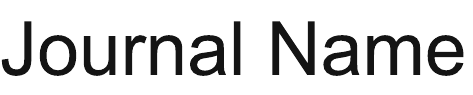}\hfill\raisebox{0pt}[0pt][0pt]{\includegraphics[height=55pt]{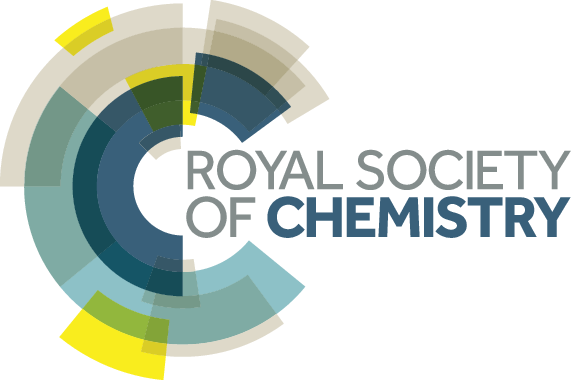}}\\[1ex]
\includegraphics[width=18.5cm]{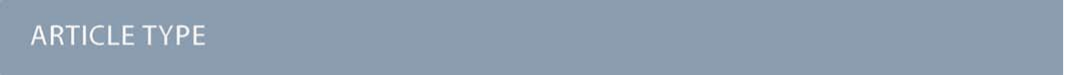}}\par
\vspace{1em}
\sffamily
\begin{tabular}{m{4.5cm} p{13.5cm} }

\includegraphics{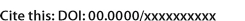} & \noindent\LARGE{\textbf{Prospect of high-temperature superconductivity in layered metal borocarbides$^\dag$}} \\
\vspace{0.3cm} & \vspace{0.3cm} \\

 & \noindent\large{Charlsey R. Tomassetti,\textit{$^{a\ddag}$} Gyanu P. Kafle,\textit{$^{a\ddag}$} Edan T. Marcial,\textit{$^{a}$} Elena R. Margine,\textit{$^{a}$} and Aleksey N. Kolmogorov$^{\ast}$\textit{$^{a}$}} \\

\includegraphics{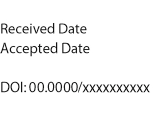} & \noindent\normalsize{Delithiation of the known layered LiBC compound was predicted to induce conventional superconductivity at liquid nitrogen temperatures but extensive experimental work over the past two decades has detected no signs of the expected superconducting transition. Using a combination of first-principles stability analysis and anisotropic Migdal-Eliashberg formalism, we have investigated possible Li$_x$BC morphologies and established what particular transformations of the planar honeycomb BC layers are detrimental to the material’s superconductivity. We propose that Li$_x$BC reintercalation with select alkali and alkaline earth metals could lead to synthesis of otherwise inaccessible metastable Li$_x$M$_y$BC superconductors with critical temperatures ($T_{\rm c}$) up to 73~K. The large-scale exploration of metal borocarbides has revealed that NaBC and Li$_{1/2}$Na$_y$BC layered phases are likely true ground states at low temperatures. The findings indicate that this compositional space may host overlooked synthesizable compounds with potential to break the $T_{\rm c}$ record for conventional superconductivity in ambient-pressure materials.} 

\end{tabular}

 \end{@twocolumnfalse} \vspace{0.6cm}

  ]

\renewcommand*\rmdefault{bch}\normalfont\upshape
\rmfamily
\section*{}
\vspace{-1cm}


\footnotetext{\textit{$^{a}$~Department of Physics, Applied Physics, and Astronomy, Binghamton University-SUNY, Binghamton, New York 13902, USA}}
\footnotetext{\textit{$^{\ast}$~E-mail: kolmogorov@binghamton.edu}}

\footnotetext{\dag~Electronic Supplementary Information (ESI) available: Additional figures and tables pertaining to stability, structural, electronic, vibrational and superconducting properties for relevant materials, x-ray diffraction comparisons with experiment, and computational details. See DOI: 00.0000/00000000.}

\footnotetext{\ddag~These authors contributed equally to this work.}



\section{Introduction}
\label{sec:introduction}

Observations of superconducting transitions in ambient-pressure MgB$_2$~\cite{Nagamatsu2001} and compressed hydrides~\cite{Drozdov2015, Drozdov2019, Bernstein2015} have shown that high-critical temperature ($T_{\rm c}$) conventional superconductivity is attainable in compounds with markedly different chemistry and morphology. The shared feature underpinning their exceptional superconducting properties is the presence of doped covalent bonds. The partially filled electronic states in these materials generate a sizable density of states (DOS) at the Fermi level while the hard vibrational modes of strongly bonded light elements are able to withstand pronounced softening from large electron-phonon (e-ph) coupling. These traits have been widely used to guide the development of new superconductors but, unfortunately, hole-doping of covalent bonds is detrimental to thermodynamic stability and the issue has been circumvented successfully in only a handful of cases. In synthesized H$_3$S and LaH$_{10}$ superhydrides with the respective $T_{\rm c}$ of 203~K~\cite{Drozdov2015} and 250~K~\cite{Drozdov2019}, formation of 3D metallized covalent frameworks at ‘unnatural’ stoichiometries is forced via application of extreme pressures. In the 55~K Q-carbon-based superconductor, the heavy B doping (27~at\%) of the predominantly $sp^3$ amorphous C network is accomplished via quenching of laser-heated samples~\cite{Bhaumik2017}. The quasi-2D MgB$_2$ with a $T_{\rm c}$ of 39~K~\cite{Nagamatsu2001} possesses naturally hole-doped B layers with a significant amount of key B-$\sigma$ states at the Fermi level because of the step-wise dependence of DOS on the number of carriers in 2D systems. The proposed closely related LiB with a similarly large electronic DOS and softened in-plane B phonon modes~\cite{ak08} forms just above 20~GPa and remains metastable at ambient conditions~\cite{ak30} but the prediction of its superconductivity up to 34~K~\cite{Kafle2022} has yet to be tested experimentally.

A prime ambient-pressure candidate is a delithiated form of the bulk LiBC material comprised of honeycomb BC layers. It was expected to supplant the record-holding MgB$_2$, with predicted $T_{\rm c}$ as high as 100~K~\cite{Rosner2002}. However, after two decades, this material has never been observed to superconduct despite extensive experimental effort~\cite{Bharathi2002, Zhao2003,Fogg2003a,Fogg2003b,Fogg2006, Kalkan2019}. The exact reasons for why Li$_x$BC ($1 > x$) has failed to live up to its full potential are still not clear. Fogg et al.~\cite{Fogg2006} sought to characterize the delithiation process and observed expulsion of B with the eventual breakdown of the quasi-2D BC network for $x$ values below $\sim 0.45$. Their density functional theory (DFT) simulations revealed the increasing prevalence of a specific defect in the BC lattice upon the reduction of Li content. A different end-product configuration was later proposed by Kalkan and Ozdas~\cite{Kalkan2019}. They inferred that, much like in graphite intercalation compounds (GICs), Li extraction from LiBC leads to the formation of Daumas-H\'{e}rold (DH)-type domains~\cite{DH}. 

As the first step toward modeling the delithiation process, our recent {\it ab initio} study focused on establishing the $(T,P_{\textrm{Li}})$ thermodynamic conditions necessary to destabilize the starting LiBC material~\cite{Kharabadze2023}. The constructed phase diagram proved to be in excellent agreement with the synthesis conditions in successful delithiation experiments~\cite{Zhao2003,Fogg2006}. Importantly, we considered all previously reported Li-B-C phases and demonstrated that representative Li$_x$BC derivatives with $1 > x\ge 0.5$ are only metastable under ambient temperature and pressure, which indicates that the covalent BC layers are strong enough to kinetically protect the delithiation pathway~\cite{Kharabadze2023}.

\begin{figure}[t!]
   \centering
\includegraphics[width=0.45\textwidth]{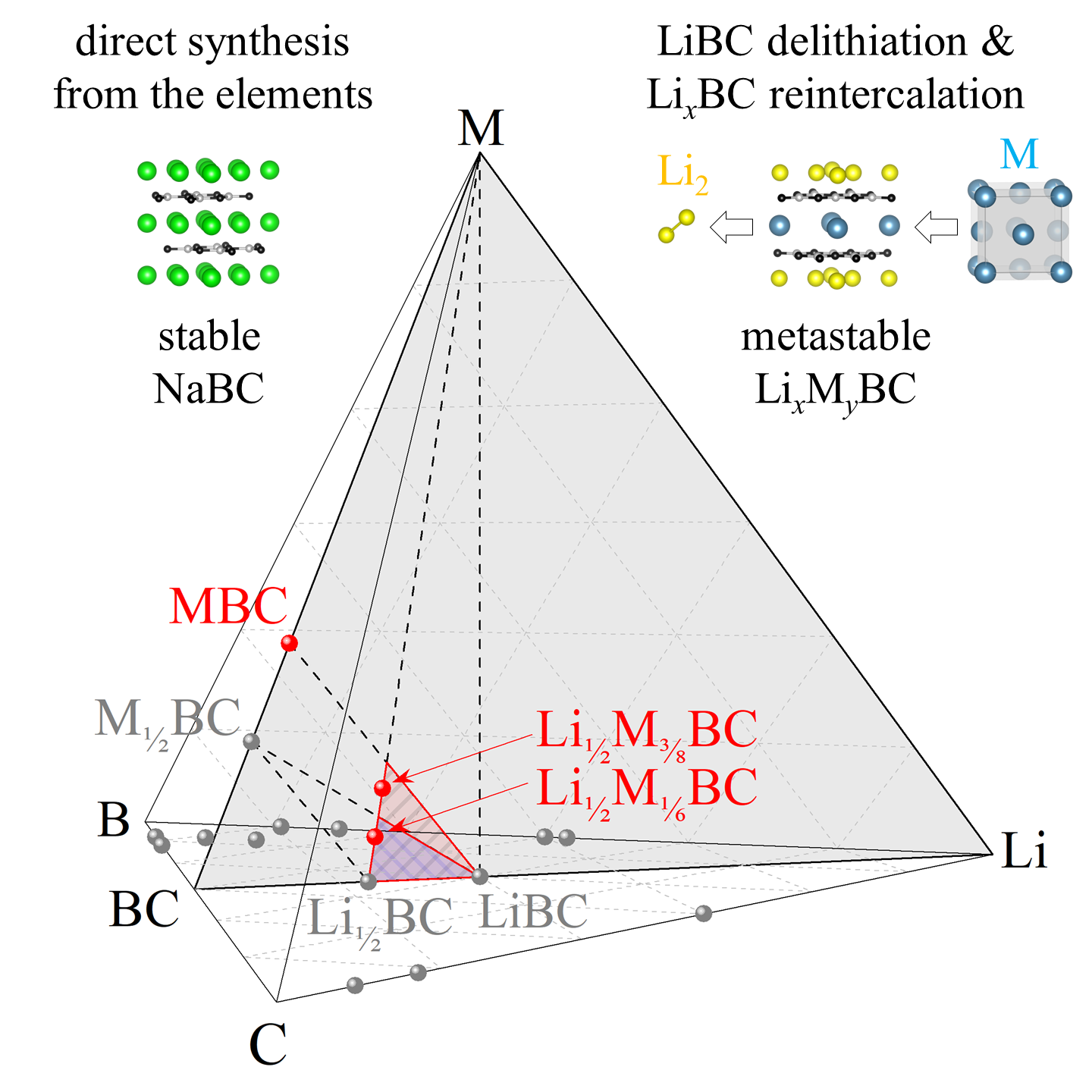}
    \vspace{-4mm}
    \caption{\label{fig-01} Compositions of observed (grey points) and some of the proposed (red points) compounds in the Li-M-B-C chemical space. The red (blue) triangles highlight the location of possible holed-doped layered Li$_x$M$_y$BC superconductors partially reintercalated with alkali (alkaline earth) metals.}
\end{figure}

In this work, we address two questions pertaining to the superconducting potential of hole-doped layered borocarbides. Firstly, we seek to attain a better understanding of the Li$_x$BC morphologies produced during experiments and use this knowledge to explain why no sample has ever been observed to superconduct. Our systematic screening of Li$_x$BC configurations uncovers a number of alternative motifs that become favored over the signature honeycomb BC framework upon the Li extraction. The following examination of the identified candidates’ properties within the anisotropic Migdal-Eliashberg (aME) formalism reveals that {\it ordered} Li$_x$BC configurations {\it should} be outstanding superconductors with $T_{\rm c}$ over 30~K but certain defects are indeed detrimental to the e-ph coupling. Secondly, we explore the possibility of obtaining synthesizable superconductors in the Li-M-BC composition subspace shown in Fig.~\ref{fig-01}. We demonstrate that reintercalation of the metastable Li$_x$BC with different metals (M = Na, Mg, K, or Ca) may be possible thermodynamically provided that the material can withstand the process in its kinetically constrained layered form. Moreover, the DFT results indicate that phases near the Li$_{1/2}$Na$_{1/2}$BC composition are tantalizingly close to the convex hull and could form from the elements. Our aME calculations show that ordered Li$_x$M$_y$BC configurations would have comparably high critical temperatures. 

The expected accessibility of these (meta)stable layered phases via reintercalation, ion exchange, electrochemical~\cite{Isaev1997, Abramchuk2017, Uppuluri2018, Bahrami2019}, or direct reactions makes the proposed materials class conceptually different from the large set of recently considered superconductors with shared chemistry in terms of synthesis conditions (e.g., at high pressures) and/or morphology (e.g., in 2D or clathrate forms): Li$_4$B$_5$C$_2$~\cite{Bazhirov2014},  Li$_3$B$_4$C$_2$ ~\cite{Gao2015}, 2D LiB$_2$C$_2$~\cite{Gao2020}, Li$_{2x}$BC$_3$~\cite{Quan2020}, M$_x$BC$_2$~\cite{Hayami2020}, MgB$_3$C$_3$~\cite{Pham2023}, 2D Mg$_2$B$_4$C$_2$~\cite{Singh2022}, X-Y-B$_2$~\cite{Wang2022}, Li-B-C~\cite{Zheng2023}, and X-Y-B$_6$C$_6$~\cite{Geng2023}. While the feasibility of obtaining our identified quaternary phases depends on a number of kinetic and thermodynamic factors that require further study, the findings dramatically expand the search space for high-$T_{\rm c}$ ambient-pressure superconductors. 

\begin{figure*}[t]
	\centering
\includegraphics[width=0.95\textwidth]{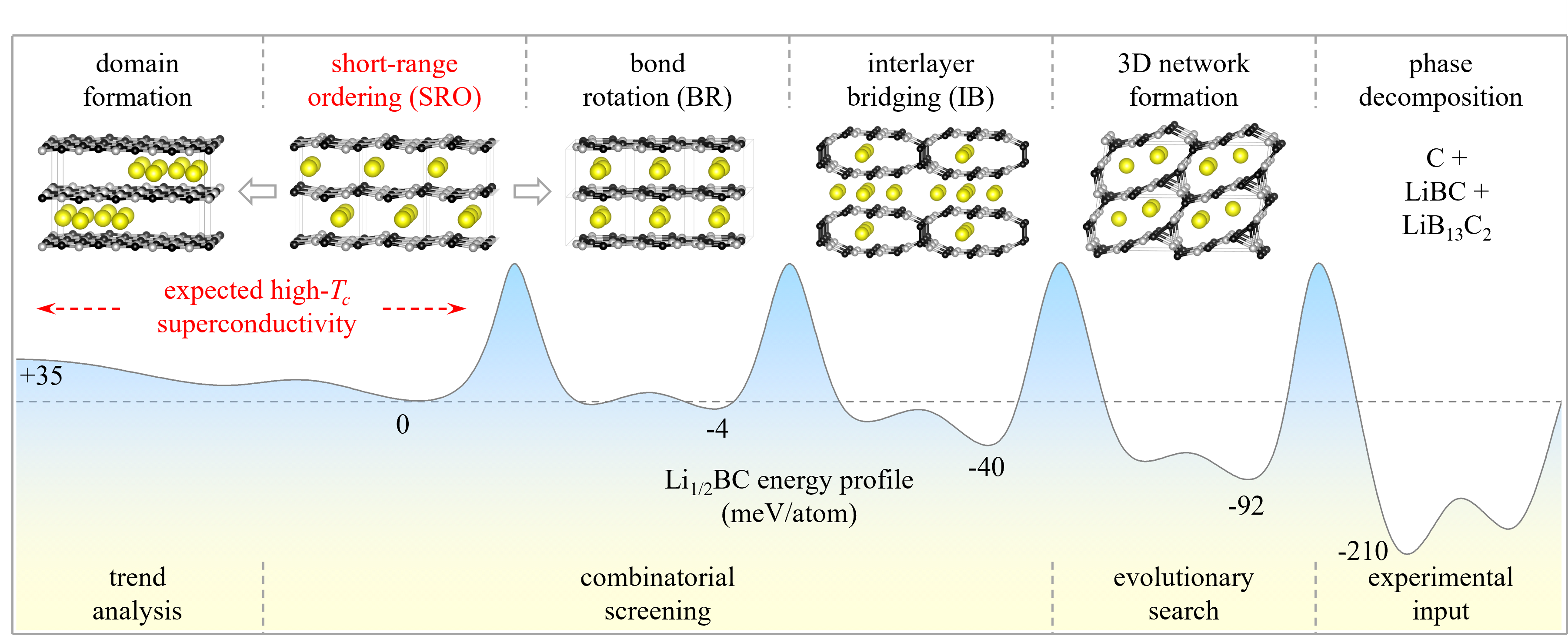}
	\caption{\label{fig-02} Schematic potential energy profile for Li$_{1/2}$BC configurations. The most favorable oI10 structure with ordered honeycomb BC layers is used as a reference and its select transformations are illustrated with representative structures. The heights of barriers separating local/global minima are not to scale. The approaches used to explore the Li$_x$BC configurational space are listed at the bottom. Our superconductivity calculations indicate that Li$_{1/2}$BC polymorphs with short- or long-range ordering of honeycomb BC layers and Li vacancies should be high-$T_{\rm c}$ superconductors.}
\end{figure*}

\section{Methods}
\label{sec:methods}

Stability analysis of Li-M-B-C phases was conducted with {\small VASP}~\cite{Kresse1996} using projector augmented wave potentials~\cite{Blochl1994} and a 500~eV plane-wave cutoff. In order to account for the dispersive interactions important in layered materials~\cite{ak06,Lebegue2010,ak30,Ning2022}, we relied on the optB86b-vdW functional~\cite{optB86b} but also used Perdew-Burke-Ernzerhof (PBE) parametrization~\cite{PBE}, optB88-vdW~\cite{optB88}, and r2SCAN+rVV10~\cite{Furness2020,Ning2022} functionals to check the sensitivity of the results to the DFT approximations. All structures were
evaluated with dense ($\Delta k \sim 2 \pi \times 0.025$~\AA$^{-1}$)  Monkhorst–Pack
$\bk$-meshes~\cite{Monkhorst1976}.

Global structure optimizations were carried with an evolutionary algorithm implemented in the MAISE package~\cite{maise}. In fixed-composition runs, randomly initialized 16-member populations with up to 21 atoms per unit cell were evolved for up to 200 generations using standard mutation and crossover operations~\cite{maise}. For the systematic screening of possible decorations of interlayer sites in metal borocarbides we started with stoichiometric MBC supercells with up to 72 atoms and sequentially substituted or/and removed metals keeping only unique configurations. Detection of similar structures for elimination of duplicates in evolutionary optimizations and combinatorial explorations was done with our fingerprint based on radial distribution functions (RDFs)~\cite{ak16,maise}. The thermodynamic corrections due to vibrational entropy were evaluated within the finite displacement methods implemented in PHONOPY~\cite{Togo2015}. We used supercells between 56 and 256 atoms, with the average of 152 atoms over 80 phases, and applied 0.1~\AA\ displacements within the harmonic approximation. According to our previous quasi-harmonic approximation results for related Li-B-C and Na-Sn materials~\cite{Kharabadze2023, thorn2023} and present tests for NaBC and Li$_{1/2}$Na$_{1/2}$BC (Figs. S15-S16), volume expansion has an insignificant effect on the formation free energies in the considered temperature range.  

Detailed information about structure, stability, and $T_{\rm c}$ for key Li-M-B-C phases discussed in the manuscript can be found in Table~\ref{table1}. To avoid ambiguity when referring to different phases at the same composition, we use Pearson symbols and space groups when needed. In addition, we indicate morphology of considered Li$_x$BC variants with superscripts using abbreviations explained in Fig.~\ref{fig-02} and below. Full structural information for relevant DFT-optimized Li-M-B-C phases is provided as CIF files. 

For calculating properties related to superconductivity, we employed the Quantum {\small ESPRESSO} package~\cite{QE} with the optB86b-vdW functional~\cite{optB86b} and norm-conserving pseudopotentials from the Pseudo Dojo library~\cite{Dojo2018} generated with the relativistic PBE parametrization~\cite{PBE}. A plane-wave cutoff value of 100~Ry, a Methfessel-Paxton smearing~\cite{Methfessel1989} value of 0.02~Ry, and $\Gamma$-centered Monkhorst-Pack~\cite{Monkhorst1976} \textbf{k}-meshes were used to describe the electronic structure. The lattice parameters and atomic positions were relaxed until the total energy was converged within $10^{-6}$~Ry and the maximum force on each atom was less than $10^{-4}$ Ry/\AA. The dynamical matrices and the linear variation of the self-consistent potential were calculated within density-functional perturbation theory~\cite{Baroni2001} on irreducible sets of regular \textbf{q}-meshes. The optimized lattice parameters for the investigated phases and the \textbf{k}- and \textbf{q}-meshes used are reported in Tables~S1 and S4~\cite{SM}. 

The EPW code~\cite{Giustino2007, EPW2016, Margine2013, EPW2023} was used to investigate the e-ph interactions and superconducting properties. The Wannier interpolation~\cite{WANN1, WANN2, WANN3} was performed on uniform $\Gamma$-centered \textbf{k}-grids (see Table~S4~\cite{SM}) using the Wannier90 code~\cite{WANN1, WANN2} in library mode. We used three $p$ orbitals for every C atom as projections for the maximally localized Wannier functions to accurately describe the electronic structure of all the compounds under investigation (see Fig.~S22~\cite{SM}). The anisotropic Migdal-Eliashberg equations were solved on fine uniform $\bk$- and $\bq$-point grids (see Table~S4~\cite{SM}), with an energy window of $\pm 0.2$~eV around the Fermi level and a Gaussian smearing value of 50~meV. The Matsubara frequency cutoff was set to 1.5~eV and we employed a uniform sampling scheme when solving the isotropic or anisotropic Migdal-Eliashberg equations. The sensitivity of $T_{\rm c}$ estimates to the $\mu^*$ parameter was assessed by performing calculations not only with the standard 0.10 value but also with 0.20.

\label{sec:Li$_x$BC}

\section{Results and Discussion}

\begin{table*}[htbp]
\centering
\caption{\label{table1} Thermodynamic stability calculated with VASP~\cite{Kresse1996} and PHONOPY~\cite{Togo2015} and superconducting critical temperature calculated with Quantum {\small ESPRESSO}~\cite{QE} and EPW~\cite{EPW2016,EPW2023} in select Li$_x$BC and Li$_x$M$_y$BC phases. The full set of observed phases used in the construction of the convex hulls is given in Fig.~S12~\cite{SM}. The phases defining the thermodynamic stability of Li$_x$BC and Li$_x$M$_y$BC phases are given as a superscript number at each temperature, and the corresponding numbered reference phases are listed as follows: (1) hP6-LiBC, (2) oF32-LiB$_6$C, (3) mS30-B$_4$C, (4) hP4-C, (5) hP6-NaBC, (6) oP14-NaB$_5$C, (7) cF4-Na, (8) cF4-K, (9) oF72-KC$_8$, (10) oS80-MgB$_2$C$_2$, (11) oI60-MgB$_{12}$C$_2$, (12) tI20-CaB$_2$C$_2$, (13) cP7-CaB$_6$, and (14) cF4-Ca. The critical temperatures were obtained by solving the isotropic Migdal-Eliashberg (iME) and anisotropic Migdal-Eliashberg (aME) with the EPW code for $\mu^*=0.10$ and $\mu^*=0.20$.} 
\setlength{\tabcolsep}{5pt}
\begin{tabular}{l l  p{0.07\textwidth} p{0.07\textwidth} p{0.07\textwidth} p{0.07\textwidth} p{0.07\textwidth} c c c c } 
\hline \hline
\noalign{\vskip 1mm} 
  Composition & Space group - &   \multicolumn{5}{c}{Distance to convex hull [meV/atom]}    & \multicolumn{4}{c}{$T_{\rm c}$ [K]}\\
      &   Pearson symbol  & \multicolumn{1}{c}{0 K}   &  \multicolumn{1}{c}{0 K}  & \multicolumn{1}{c}{300 K}  & \multicolumn{1}{c}{600 K}  &  \multicolumn{1}{c}{900 K}    & iME  & iME  & aME  & aME   \\ 
      &       &    &  \multicolumn{1}{c}{(+ZPE)}  &   &   &    & 0.10 & 0.20 & 0.10 &  0.20  \\ \hline \noalign{\vskip 1mm}
      Li$_{1/3}$BC$^{\rm 3D}$ & $Amm2$-oS14     & 155.7$^{1,2,4}$& & & & & & & &\\  \noalign{\vskip 1mm}  
         Li$_{1/2}$BC$^{\rm 3D}$ &  $C/m$-mS20  & 118.2$^{1,2,4}$ & 111.4$^{1,3,4}$ & 111.0$^{1,3,4}$  & 109.3$^{1,3,4}$ &  107.0$^{1,3,4}$ &  & ~ & ~ & ~ \\  \noalign{\vskip 1mm} \hline \noalign{\vskip 1mm} 
       Li$_{1/2}$BC$^{\rm SRO}$ & $Pnnm$-oP10               & 210.6$^{1,2,4}$ & 195.1$^{1,3,4}$ & 191.5$^{1,3,4}$ & 182.8$^{1,3,4}$ & 172.6$^{1,3,4}$ & 17.4 & 9.8 & 42 & 32 \\  \noalign{\vskip 1mm} 
       Li$_{1/2}$BC$^{\rm SRO}$ & $Imm2$-oI10            & 209.9$^{1,2,4}$ & 195.0$^{1,3,4}$ & 191.8$^{1,3,4}$ & 184.1$^{1,3,4}$ & 174.9$^{1,3,4}$ & 33.7 & 23.0 & 49 & 38 \\  \noalign{\vskip 1mm}  
       Li$_{1/2}$BC$^{\rm RH}$ & $P\bar{3}m1$-hP5        & 244.6$^{1,2,4}$ & 229.5$^{1,3,4}$ & 227.5$^{1,3,4}$ & 221.7$^{1,3,4}$ & 214.4$^{1,3,4}$ & 49.8 & 36.7 & 84 & 71 \\  \noalign{\vskip 1mm}  
       Li$_{1/2}$BC$^{\rm SRO}$ & $Pmma$-oP10             & 220.9$^{1,2,4}$ & ~ & ~ & ~ & ~                                                        & 33.9 & 23.1 & 82 & 70 \\  \noalign{\vskip 1mm} 
       Li$_{1/2}$BC$^{\rm SRO}$ & $P\bar{3}m1$-hP15                   & 225.5$^{1,2,4}$ & 209.1$^{1,3,4}$ & 205.5$^{1,3,4}$ & 196.7$^{1,3,4}$ & 186.2$^{1,3,4}$ & 53.3 & 40.2 & 72 & 60 \\  \noalign{\vskip 1mm}   
       Li$_{1/2}$BC$^{\rm SRO}$ & $P312$-hP15                   & 219.5$^{1,2,4}$ &  &  &  &  & &  &  &  \\  \noalign{\vskip 1mm} \hline \noalign{\vskip 1mm}  
        Li$_{1/2}$BC$^{\rm IB}$ & $P2/m$-mP20   & 169.9$^{1,2,4}$ & 159.6$^{1,3,4}$ & 157.4$^{1,3,4}$ & 152.0$^{1,3,4}$ & 145.6$^{1,3,4}$ & 0.5 & ~ & ~ & ~ \\  \noalign{\vskip 1mm} 
        Li$_{1/2}$BC$^{\rm IB}$ & $Pmm2$-oP20   & 186.9$^{1,2,4}$ & ~ & ~ & ~ & ~                                                       & 1.6 & ~ & ~ & ~ \\  \noalign{\vskip 1mm} 
        Li$_{1/2}$BC$^{\rm BR}$ & $P/m$-mP15    & 205.8$^{1,2,4}$ & 197.4$^{1,3,4}$ & 194.1$^{1,3,4}$ & 186.8$^{1,3,4}$ & 178.9$^{1,3,4}$ & 0.0 & ~ & ~ & ~ \\  \noalign{\vskip 1mm}  
        Li$_{1/2}$BC$^{\rm BR}$ &  $P/m$-mP30  & 231.4$^{1,2,4}$ &  &  &  &  &  & ~ & ~ & ~ \\  \noalign{\vskip 1mm}  
        Li$_{1/2}$BC$^{\rm BR}$ &  $P1$-aP15  & 208.1$^{1,2,4}$ &  &  &  &  &  & ~ & ~ & ~ \\  \noalign{\vskip 1mm} \hline \noalign{\vskip 1mm} 
        Li$_{5/8}$BC$^{\rm SRO}$ & $C2/m$-mS42               & 148.3$^{1,2,4}$ & ~ & ~ & ~ & ~                                                    & 40.5 & 28.4 & 59 & 48 \\  \noalign{\vskip 1mm} 
        Li$_{2/3}$BC$^{\rm SRO}$ & $P6_3/mcm$-hP16            & 118.0$^{1,2,4}$ & ~ & ~ & ~ & ~                                                   & 47.0 & 34.0 & 83 & 71 \\  \noalign{\vskip 1mm} 
        Li$_{3/4}$BC$^{\rm SRO}$ & $Pmma$-oP22               & 85.9$^{1,2,4}$ & 78.9$^{1,3,4}$ & 77.8$^{1,3,4}$ & 74.7$^{1,3,4}$ & 71.0$^{1,3,4}$ & 42.6 & 29.6 & 82 & 72 \\  \noalign{\vskip 1mm} 
        Li$_{5/6}$BC$^{\rm SRO}$ & $P\bar{3}1m$-hP17          & 55.6$^{1,2,4}$ & ~ & ~ & ~ & ~                                                    & 28.6 & 17.3 & 59 & 49 \\  \noalign{\vskip 1mm} \hline \noalign{\vskip 1mm} 
        Li$_{1/2}$Na$_{1/2}$BC &  $P\bar{3}m1$-hP6  & 3.5$^{1,5}$ & 3.5$^{1,5}$ & 3.6$^{1,5}$ & 11.4$^{1,4,6,7}$ & 22.2$^{1,4,6,7}$ &  & ~ & ~ & ~ \\  \noalign{\vskip 1mm} 
        Li$_{1/2}$Na$_{5/12}$BC &  $P2/m$-mP35       & 6.9$^{1,4,5,6}$ & 3.3$^{1,4,5,6}$ & 2.2$^{1,4,5,6}$ & 6.3$^{1,4,6,7}$ & 12.4$^{1,4,6,7}$ & ~ & ~ & ~ & ~ \\  \noalign{\vskip 1mm} 
        Li$_{1/2}$Na$_{3/8}$BC & $P2/m$-mP23        & 9.6$^{1,4,5,6}$ & 5.0$^{1,4,5,6}$ & 3.5$^{1,4,5,6}$ & 6.0$^{1,4,6,7}$ & 10.2$^{1,4,6,7}$      & 28.5 & 17.1 & 75 & 67 \\  \noalign{\vskip 1mm} 
        Li$_{1/2}$Na$_{1/3}$BC & $P\bar{3}m1$-hP17 & 20.8$^{1,4,5,6}$ & 16.0$^{1,4,5,6}$ & 15.2$^{1,4,5,6}$ & 17.9$^{1,4,6,7}$ & 21.9$^{1,4,6,7}$ & 35.5 & 23.1 & 77 & 63 \\  \noalign{\vskip 1mm} 
        Li$_{1/2}$Na$_{1/3}$BC &  $P2/c$-mP34      & 15.4$^{1,4,5,6}$ & 10.8$^{1,4,5,6}$&  9.0$^{1,4,5,6}$ & 10.5$^{1,4,6,7}$ & 13.3$^{1,4,6,7}$ &  & ~ & ~ & ~ \\  \noalign{\vskip 1mm} 
        Li$_{1/2}$Na$_{1/4}$BC & $P2/m$-mP11      & 56.7$^{1,4,5,6}$ & ~ & ~ & ~ & ~                                                             & 32.9 & 21.6 & 74 & 62 \\  \noalign{\vskip 1mm} 
        Li$_{1/2}$K$_{1/4}$BC & $P2/c$-mP22  & 35.6$^{1,2,8,9}$ & ~ & ~ & ~ & ~                                                             &  &  &  &  \\  \noalign{\vskip 1mm} 
        Li$_{1/2}$K$_{1/6}$BC & $P\bar{3}1m$-hP16  & 79.5$^{1,2,8,9}$ & ~ & ~ & ~ & ~                                                             & 43.7 & 31.5 & 84 & 73 \\  \noalign{\vskip 1mm} 
        Li$_{1/2}$Mg$_{1/6}$BC & $P\bar{3}1m$-hP16 & 61.4$^{1,4,10,11}$  & ~ & ~ & ~ & ~                                                          & 33.7 & 21.7 & 62 & 49 \\  \noalign{\vskip 1mm} 
        Li$_{1/2}$Ca$_{1/6}$BC & $P\bar{3}m1$-hP16 & 57.7$^{1,4,12,13}$ & ~ & ~ & ~ & ~                                                           & 34.8 & 23.1 & 58 & 45 \\ \noalign{\vskip 1mm} 
        Li$_{1/2}$Ca$_{1/3}$BC &  $P\bar{3}m1$-hP17 & 74.8$^{1,12,14}$ & ~ & ~ & ~ & ~                                                                             & 2.4 &  &  \\ \noalign{\vskip 1mm} 
        Li$_{1/2}$Ca$_{1/2}$BC &  $P\bar{3}m1$-hP6 & 120.0$^{1,12,14}$ & ~ & ~ & ~ & ~                                                                             & 1.8 &  &  \\ \noalign{\vskip 1mm} 
        
 \noalign{\vskip 1mm}  \hline  
 \noalign{\vskip 1.5mm} 
\label{table1}
\end{tabular}
\end{table*}

\subsection{Background of \texorpdfstring{Li$_x$BC}{Lg}}

After Rosner et al. theorized that hole-doped lithium borocarbides could exhibit superconductivity up to 100~K~\cite{Rosner2002}, several groups have synthesized delithiated LiBC~\cite{Bharathi2002, Fogg2003a, Fogg2003b, Zhao2003, Fogg2006, Kalkan2019} and concluded from resistivity measurements that the material shows no sign of the expected transition down to 2~K. The studies found that it is difficult to control the exact Li content removed due to the dynamic and nonequilibrium nature of the delithiation reaction. The typical route for obtaining Li$_x$BC involved high temperature Li deintercalation~\cite{Fogg2003a, Zhao2003, Fogg2006, Kalkan2019}. At the higher 1500$^{\circ}$C end of the investigated temperature range, experiments produced two groups of phases, one with Li contents $x>0.8$ and the other with $0.6>x$~\cite{Fogg2003a, Fogg2006, Kalkan2019}. The use of low-pressure setups allowed Zhao et al.~\cite{Zhao2003} to induce deintercalation at a lower 600-800$^{\circ}$C temperature and produce three hole-doped samples of nominal compositions Li$_{0.80}$BC, Li$_{0.77}$BC, and Li$_{0.63}$BC. The use of the oxidizing agent NOBF$_4$ in acetonitrile made it possible to extract Li at near-ambient 95$^{\circ}$C but Raman spectra indicated that samples with $x=0.5$ and 0.65 consisted of a disordered graphite-like BC material alongside unreacted LiBC~\cite{Fogg2003b}.

The study by Fogg et al.~\cite{Fogg2006} took a particular focus on the delithiation limit of Li$_x$BC. After observing the expulsion of B-rich byproducts from their Li-poor ($0.45>x$) samples, they conjectured that a type of defect consisting of a B-C atom swap could happen at higher compositions ($x>0.45$). The increasing occurrence of this defect with decreasing Li content was suggested to kinetically ease the degeneration of Li$_{0.45>x}$BC into non-layered C-rich and B-rich phases. Their DFT calculations agreed that such a transformation would become energetically favorable at Li content $x=0.5$, with recent work~\cite{Kharabadze2023} corroborating that a low-symmetry structure with swapped B and C atoms becomes slightly favored over the standard morphology with perfectly alternating B-C sites. Bond rotation is a commonly observed transformation in covalent honeycomb systems, such as the dislocation dynamics of the 5-7 Stone-Wales defect pairs in graphene probed with low-voltage high-resolution transmission electron microscopy~\cite{warner2012} or the appearance of B-B bonds in Li$_x$BC, indicative of B-C swaps, detected with spectroscopic methods in the discharged LiBC anode material tested recently for Li-ion battery applications~\cite{Peng2023}.

The ability of borocarbides to form graphitic honeycomb networks raises the question of whether they exhibit properties similar to that of GICs~\cite{Dresselhaus2002} or transition metal dichalcogenides (TMDs)~\cite{Friend1987}, including stage formation with a DH-type domain intercalation mechanism~\cite{DH}. Staging is the phenomenon in which intercalants are placed between $n$ layers of unintercalated host material, with said material classified as stage-$n$. 
At low stages, the domain size is expected to be large~\cite{Dresselhaus2002}, which does not preclude the possibility of mixed-stage samples and staging disorder. Based on the observation of a 12$^\circ$ (001) powder x-ray diffraction (PXRD) peak in Li$_{0.56}$BC, Kalkan and Ozdas~\cite{Kalkan2019} surmised the existence of stage-2 ordering, known to occur in GICs~\cite{Dresselhaus2002} and TMDs~\cite{Friend1987} but not yet detected in Li$_x$BC, as other studies reported reflections only down to 15-20$^\circ$~\cite{Nakamori2003, Fogg2003b, Zhao2003}.
Kalkan and Ozdas further proposed a mixed-stage phase of 24\% (wt \%) stage-1 LiBC and 76\% stage-2 Li$_{0.43}$BC as the best fitting model to the PXRD data. Moreover, their Reitveld analysis offers evidence of shifting and buckling of the BC layers, similar to that observed in Mg$_{1/2}$BC~\cite{worle1994}.

\subsection{Stability of \texorpdfstring{Li$_x$BC}{Lg} }

The investigation of the Li$_x$BC subspace in our previous study of the full Li-B-C ternary focused on finding (i) the lowest-energy partially delithiated configurations with ordered honeycomb BC layers in the $3/4 \ge x \ge 1/4$ range by screening LiBC supercells with up to 18 atoms and (ii) alternative stable morphologies with $1/2 \ge x$ by performing unconstrained evolutionary searches~\cite{Kharabadze2023}. The present thermodynamic stability analysis has been extended to identify a wider variety of metastable Li$_x$BC structures relevant in the context of kinetics-defined delithiation experiments and examine their favorability across the composition range. The exploration of this configuration space involved a suite of complementary strategies detailed below and in the Methods Section.

\begin{figure}[t!]
   \centering
\includegraphics[width=0.48\textwidth]{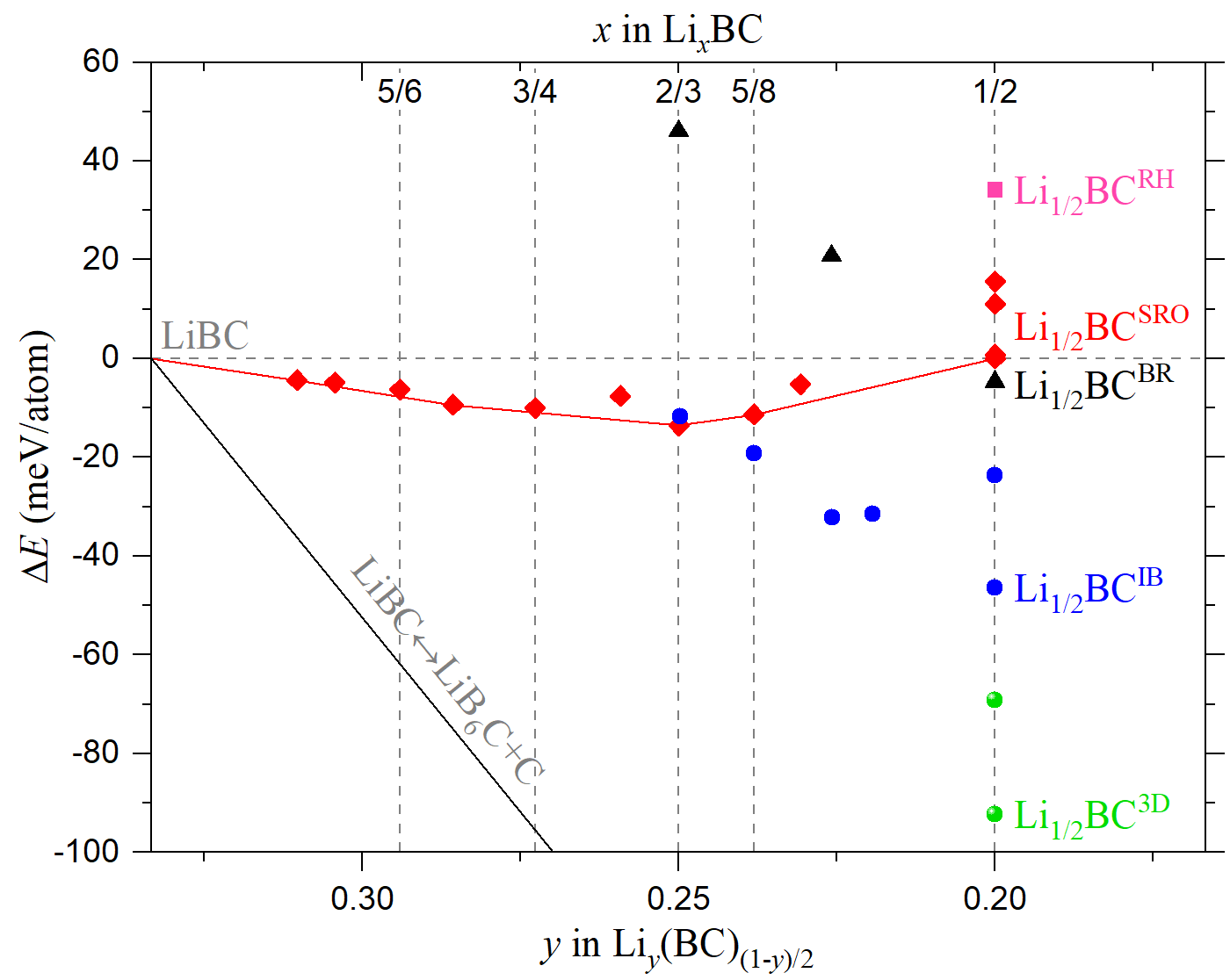}
    \caption{\label{fig-03} Relative energies of Li$_x$BC phases referenced to LiBC and the lowest-energy oI10-Li$_{1/2}$BC$^{\rm SRO}$. The notation of the examined morphologies, shown in different colors, is explained in Figs.~\ref{fig-02} and~\ref{fig-04}. The gray solid line represents the boundary of the convex hull. The lowest-energy Li$_{1/2}$BC structures for each morphology type are hP5$^{\rm RH}$, oI10$^{\rm SRO}$, mP15$^{\rm BR}$, mP20$^{\rm IB}$, and mS20$^{\rm 3D}$ (see Table~\ref{table1})}.
\end{figure}

First, we carried out a broader set of global structure searches using the evolutionary algorithm implemented in the MAISE package~\cite{maise}. We (re)examined key stoichiometries around $x=1/2$, namely, 2/3, 5/8, 1/2, 3/8, 1/3, etc., to establish when non-honeycomb motifs become favored in the deintercalation process. Previously, we observed that Li depletion below $x=1/3$ triggers the BC layers to fuse into fully-connected 3D frameworks producing, e.g., oS14-Li$_{1/3}$BC$^{\rm 3D}$ 120~meV/atom below the LiBC$\leftrightarrow$BC tie-line~\cite{Kharabadze2023}. The present evolutionary optimizations of 16-member populations with up to 21 atoms per unit cell for up to 200 generations, totaling 30,000 local optimizations across all compositions, helped us refine these findings. We have now detected emergence of more stable 3D BC frameworks at a higher $x=1/2$ Li concentration. For instance, the new mS20-Li$_{1/2}$BC$^{\rm {3D}}$ phase with a porous BC scaffold hosting double rows of Li in 1D channels (Fig.~\ref{fig-02}), was found to be nearly 90~meV/atom lower in energy (Fig.~\ref{fig-03}) than any half-filled layered variant considered previously~\cite{Fogg2006,Kharabadze2023}. Inspection of local minima produced in the evolutionary searches revealed that bridging between two layers through vacant Li sites is favorable enough to occur at $x>1/2$. The large unit cell sizes needed to represent such phases with nontrivial compositions, e.g., 5:8:8 or 7:8:8, made it difficult to ensure that the candidates found ‘from scratch’ were true ground states.

To pinpoint the crossover composition between the ordered and altered layered morphologies, we proceeded with a combinatorial screening of intercalant decorations in known and identified BC frameworks. The procedure involved generating different LiBC (super)cells and sequentially removing the metal atoms while keeping only non-equivalent configurations. Duplicate structures were identified using our structural fingerprint based on the radial distribution function~\cite{ak16,maise}. We focused on the $1\ge x \ge 1/2$ range in Li$_x$BC structures with three competing BC patterns: the original honeycomb layers displaying short-range order (SRO), interlayer bridging (IB) between C atoms, and defective layers comprising B-C bond rotations (BR).

Starting with perfectly ordered honeycomb BC layers, we created various supercells to examine both AA’ and AA stacking variations corresponding to the BC-CB and BB-CC placements of the B and C atoms along the $c$ axis, respectively. Overall, we generated 12 supercells with up to 54 atoms and ultimately evaluated over 2,500 distinct structures. At the $x=1/2$ composition, the more extensive screening reproduced the previously identified oI10-Li$_{1/2}$BC$^{\rm SRO}$ phase with buckled layers as the most stable one with this morphology. At two other previously sampled $x>1/2$ stoichiometries, the new best hP16-Li$_{2/3}$BC and oP22-Li$_{3/4}$BC represented only minor improvements in stability, by less than 1~meV/atom. The sampling of additional $x>1/2$ fractions revealed no particularly favorable ordered configurations, with the cluster of red points in Fig.~\ref{fig-03} reaching a modest minimum of -13.6~meV/atom below the LiBC$\leftrightarrow$Li$_{1/2}$BC tie-line at $x=2/3$. 

Interestingly, for some Li vacancy arrangements at the lowest considered filling fraction of $x=1/2$, the interlayer bridging occurred naturally upon simple local relaxation of the starting planar geometries. To investigate the possibility of bridging at higher Li contents, we prepared supercells with up to 48 atoms extended along one in-plane direction, removed two neighboring Li atoms, brought two C atoms in adjacent layers half-way to within the covalent distance across the vacant metal sites, and systematically probed possible arrangements of remaining Li atoms. After local optimization of over 600 candidate phases, we established that the IB configurations indeed become preferred over the SRO phases below $x=2/3$. The increasing predominance of the motif, manifested in the relative energy drop from $-12$~meV/atom at $x=2/3$ down to $-46$~meV/atom by $x=1/2$ in Fig.~\ref{fig-03}, precipitates the eventual linkage of layers into fully connected 3D frameworks.

\begin{figure*}[t!]
   \centering
\includegraphics[width=0.99\textwidth]{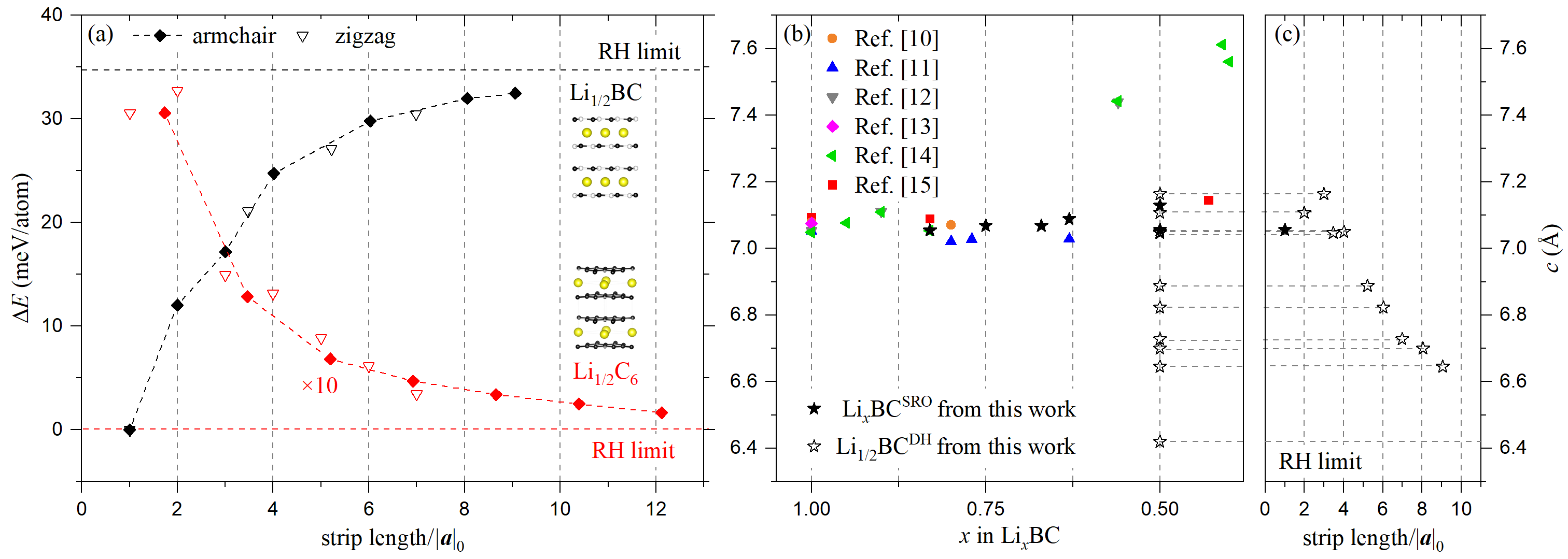}
    \caption{\label{fig-04} Stability and structural parameters of Li-intercalated materials. (a) Relative energies of Li$_{1/2}$BC (Li$_{1/2}$C$_{6}$) phases referenced to the lowest-energy oI10-Li$_{1/2}$BC$^{\rm SRO}$ (R{\"u}dorff-Hofmann stage-2 Li$_{1/2}$C$_{6}$) as a function of the domain length in the Daumas-H{\'e}rold model in units of the $a$ lattice constant, with the armchair (zigzag) sets corresponding to DH boundaries along (perpendicular to) the covalent bonds. (b) The composition-dependent $c$ lattice constants in Li$_x$BC models fully relaxed in our DFT calculations (black stars) and reported in experimental studies (other symbols). (c) $c$ lattice constant in Li$_{1/2}$BC as a function of the domain length. }
\end{figure*}

The sampling of a larger set of unit cells with rotated BC bonds improved on the aP15-Li$_{1/2}$BC$^{\rm {BR}}$ phase found previously with the evolutionary search. The new mP15$^{\rm {BR}}$ configuration at this composition is slightly more stable by 2.3 and 4.1~meV/atom compared to the aP15$^{\rm {BR}}$ and oI10$^{\rm {SRO}}$ variants, respectively. As was discussed previously~\cite{Fogg2006,Kharabadze2023}, the likelihood of this defect type is heavily suppressed at compositions above $x=1/2$. Our screening of unit cells with higher Li content revealed that the most stable structure with rotated bonds at $x=2/3$ was disfavored by nearly 50~meV/atom (Fig.~\ref{fig-03}).

Finally, we performed a comparative analysis of the stability and structural trends in the half-filled borocarbide (Li$_{1/2}$BC) and graphite (Li$_{1/2}$C$_{6}$) intercalation compounds. It should first be noted that stage-2 hP13-Li$_{1/2}$C$_{6}$ is a thermodynamically stable phase in the Li-C binary, while stage-2 hP5-Li$_{1/2}$BC is nearly 35~meV/atom higher in energy than our most stable oI10-Li$_{1/2}$BC$^{\rm {SRO}}$. In the DH domain model, intercalant islands left over from the delithiation process define boundaries between different staging domains. To simulate the energetics of increasingly large islands, we adopted a strip model with edges along either armchair or zigzag orientations of the honeycomb BC lattice. A representative armchair unit cell of length $a=10.80$~\AA\ is shown in Fig.~\ref{fig-02}. The resulting relative energies in both compounds are plotted in Fig.~\ref{fig-04}(a) as a function of the domain length. As the strip size increases, the energy gradually converges towards a R{\"u}dorff-Hofmann-type (RH) model~\cite{RH}, a perfectly boundaryless stage-2 configuration. These findings illustrate a stark difference between the energetics of domain formation in the Li-intercalated graphite and borocarbide compounds and disagree with the proposed explanation of the stage-2 configuration emerging in the latter around $x=1/2$ ~\cite{Kalkan2019}. 

Further evidence against the DH/RH models in Li$_x$BC can be seen in the material’s structural response to the Li island formation summarized in Fig.~\ref{fig-04}(b-c). The $c$ lattice constants extracted from XRD or neutron diffraction measurements~\cite{Bharathi2002, Zhao2003, Fogg2003a, Fogg2003b, Fogg2006, Kalkan2019} show little variation during delithiation and remain between 7.0~\AA\ and 7.1~\AA\ down to $x\approx 0.6$. At lower Li concentrations, the reported values in samples obtained via different synthesis routes disperse over a significant range, from 7.14~\AA\ at $x=0.43$~\cite{Kalkan2019} to 7.44~\AA\ at $x=0.56$~\cite{Fogg2003a,Fogg2006} and up to 7.61~\AA\ at $x\approx 0.4$~\cite{Fogg2006}. Our calculated $c$-axis lattice constants agree well with the experimental values for the $1\ge x  \gtrsim 0.6$ compositions, which indicates a good performance of the chosen van der Waals (vdW) functional for the description of the interlayer interactions in Li$_x$BC. At $x=1/2$, the DFT results demonstrate that the $c$ lattice constant would remain around 7.1~\AA\ if the products maintained the SRO configurations but would decrease dramatically, down to nearly 6.4~\AA, if the compound indeed adopted the stage-2 morphology. The trend differs from the well-established expansion of the interlayer distance in the Li-intercalated graphite en route from LiC$_6$ (measured 3.687 \AA~\cite{Vadlamani2014} and our calculated 3.603 \AA) to stage-2 Li$_{1/2}$C$_6$ (measured 7.0229~\AA~\cite{Vadlamani2014} and our calculated 6.936~\AA).

Therefore, staging does not seem to be a likely scenario behind the appearance of the key (001) XRD peak in Li$_x$BC samples obtained in Kalkan and Ozdas’ experiments~\cite{Kalkan2019}, and other possibilities may need to be considered to explain the evident doubling of the period along the $c$-axis. We have simulated XRD patterns for a variety of identified competing Li$_x$BC configurations and plotted them in Fig. S1~\cite{SM}. It is clear that structures with broken hexagonal symmetry caused by layer buckling, bond rotation, or interlayer bridging cannot fit the observed signature peaks. The closest match is our hP15-Li$_{1/2}$BC$^{\textrm {SRO}}$ ($P312$) phase, 9.6~meV/atom above oI10$^{\textrm {SRO}}$, which is a $\sqrt{3}\times\sqrt{3}$ supercell with 1/3 and 2/3 ordered Li populations in adjacent galleries (see Supporting Information~\cite{SM}). In fact, all considered $\sqrt{3}\times\sqrt{3}$ supercells for $5/6\ge x\ge 1/2$ produce well-matching reflection patterns except for the weak (100) peak around 22$^{\circ}$ that is not present in the collected data (the peak at the nearby 21$^{\circ}$ angle was attributed to the LiB$_3$ delithiation byproduct~\cite{Kalkan2019}).

Overall, our thermodynamic stability analysis indicates that Li$_x$BC prefers to retain perfectly ordered honeycomb layers down to about $x=2/3$ and then shows propensity to forming interlayer bridges that ultimately link BC layers into extended 3D frameworks once the Li content reduces to $x=1/2$. The periodic structures with rotated bonds become only marginally preferred over the SRO variants at $x=1/2$. In fact, the results in Table~\ref{table1} show that the inclusion of zero point energy alone reverses the order in favor of the latter. The vibrational entropy contributions bring the full set of the considered morphology types at $x=1/2$ closer to the convex hull boundary by 24-39 meV/atom at $T=900$~K but neither make them globally stable and nor cause any other significant changes in the relative ranking. The configurational entropy, estimated in our previous study to reach 22~meV/atom at 900~K~\cite{Kharabadze2023}, may promote disordered decorations of Li sites but delithiation at lower temperatures and/or subsequent annealing are expected to counteract the disorder of Li$_x$BC structures. 

Given the variety of experimental conditions and the significant challenges of simulating large-scale kinetic processes, such as staging or solid state transformations involving covalent rebonding, it is difficult to establish what particular metastable phases are produced in specific experiments. Nevertheless, the library of possible Li$_x$BC phases gives us a chance to probe the response of superconducting properties to a variety of factors.

\subsection{Superconductivity of \texorpdfstring{Li$_x$BC}{Lg}}

The primary purpose of the following analysis is to establish how possible compositional and morphological transformations taking place in the delithiation process affect the Li$_x$BC derivatives’ superconducting properties. It has been widely discussed that the starting LiBC compound is isoelectronic and isostructural to MgB$_2$~\cite{Rosner2002, Dewhurst2003} but, being a semiconductor, it has no signature hole-doped $\sigma$ states at the Fermi level essential for the high-$T_{\rm c}$ superconductivity. In the delithiated form, the material does develop multiple Fermi surface (FS) sheets of quasi-2D $p$-$\sigma$ or 3D $p$-$\pi$ character shown in Fig.~S2~\cite{SM}. So far, Li$_x$BC superconducting properties have been investigated only in a handful studies with at best the isotropic Eliashberg calculations~\cite{An2002, Dewhurst2003}. Considering the importance of anisotropic effects in the formation of multiple superconducting gaps that define the critical temperature in related layered conventional superconductors, e.g., MgB$_2$~\cite{Choi2002,Floris2007,Kafle2022}, GICs~\cite{Sanna2012, Margine2016}, and Li-Mg-B~\cite{Kafle2022}, we have relied on the aME formalism to reexamine the e-ph coupling in Li$_x$BC as well.

\begin{figure*}[t]
	\centering
    \includegraphics[width=0.99\linewidth]{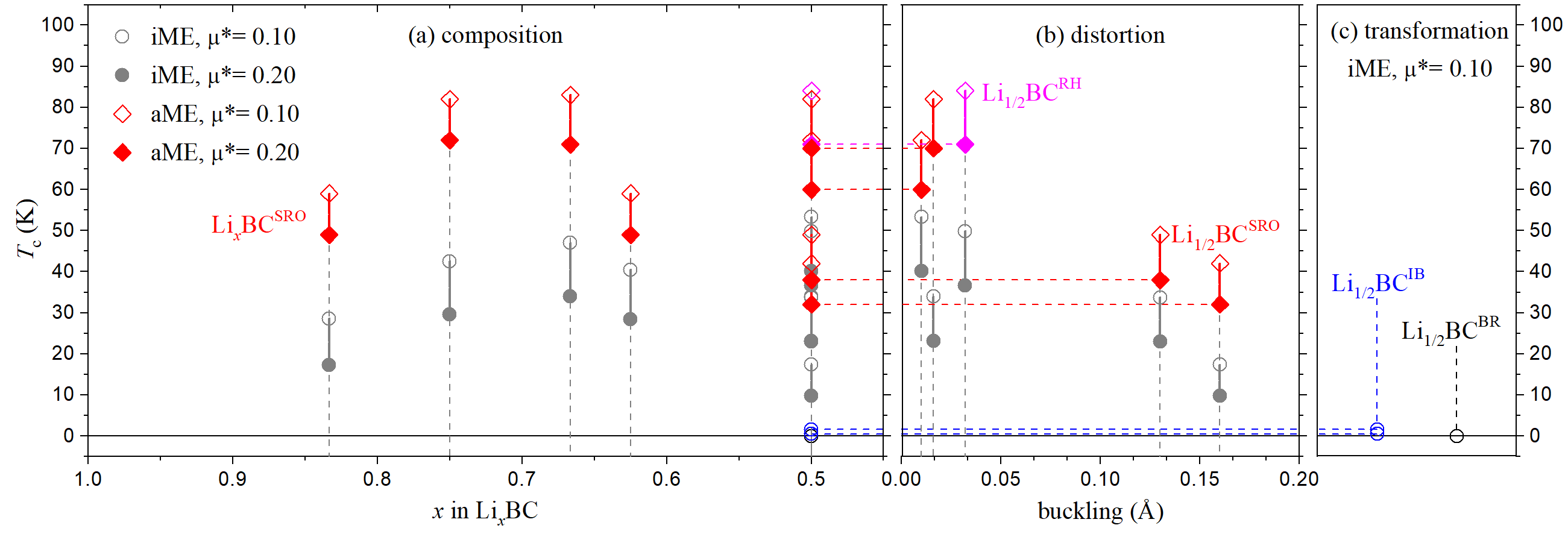}
	\caption{\label{fig-06} Dependence of critical temperature in Li$_x$BC phases on the (a) composition, (b) degree of BC layer distortion, and (c) BC layer transformation. Diamond (circle) symbols denote values obtained with the anisotropic (isotropic) ME formalism. Panel (c) shows that interlayer bridging and bond rotation effectively suppress superconductivity in Li$_{1/2}$BC.}
\end{figure*}

The $T_{\rm c}$ dependence on the Li concentration in Li$_x$BC has been previously considered only within the jellium model~\cite{Dewhurst2003}. The study found a consistent boost in the isotropic e-ph coupling strength ($\lambda$) with the increasing hole concentration, triggered by a rise in the DOS at the Fermi level, $N(E_{\textrm{F}}$), and softening of the logarithmic average frequency. In the present work, we have performed calculations for a series of the most stable SRO configurations identified across the $5/6\ge x\ge 1/2$ compositions. Their representation with ordered unit cells allows one to probe effects beyond the rigid band approximation arising not only from a non-uniform charge distribution but also from variable shifts of different $s$ and $p$ states on B/C atoms around Li vacancies, which proved to be important for establishing the Li$_x$B optimal stoichiometry~\cite{ak09} and evaluating the LiB DOS response to doping~\cite{ak10}. Table~S1~\cite{SM} shows a slightly upward trend for the important $N_{\sigma}(E_{\textrm{F}}$) with the number of holes but the total DOS at the Fermi level exhibits a non-monotonous behavior. The analysis of the phonon spectra (Fig.~S3~\cite{SM}) reveals that the frequencies of the predominate large-coupling bond-stretching modes at $\Gamma$ harden from about 80~meV ($x=5/6$) to about 90~meV ($x=1/2$) as $x$ decreases, which coincides with the steady increase of the e-ph coupling strength from $0.7$ to 1.0 (see Fig.~S3~\cite{SM}). However, the aME results indicate that superconducting properties are sensitive to both composition and structure, as the $T_{\rm c}$ in SRO phases with $5/6\ge x>1/2$ scatter between 48~K and 72~K (at $\mu^*=$0.20, see Fig.~S4~\cite{SM}) and even more so for various polymorphs at $x=1/2$, as shown in Fig.~\ref{fig-06}(a,b). Note that in addition to the standard $\mu^*$ value of 0.10~\cite{Pellegrini2022}, we used a value of 0.20 for all considered compounds to assess the sensitivity of $T_{\rm c}$ on this semi-empirical parameter. The latter provides more conservative estimates, lower by about 10~K (20\%), for MgB$_2$-type superconductors~\cite{Golobulov2002,Margine2013,Kafle2022} (see Table~\ref{table1}). 

\begin{figure*}[t]
	\centering
    \includegraphics[width=0.99\textwidth]{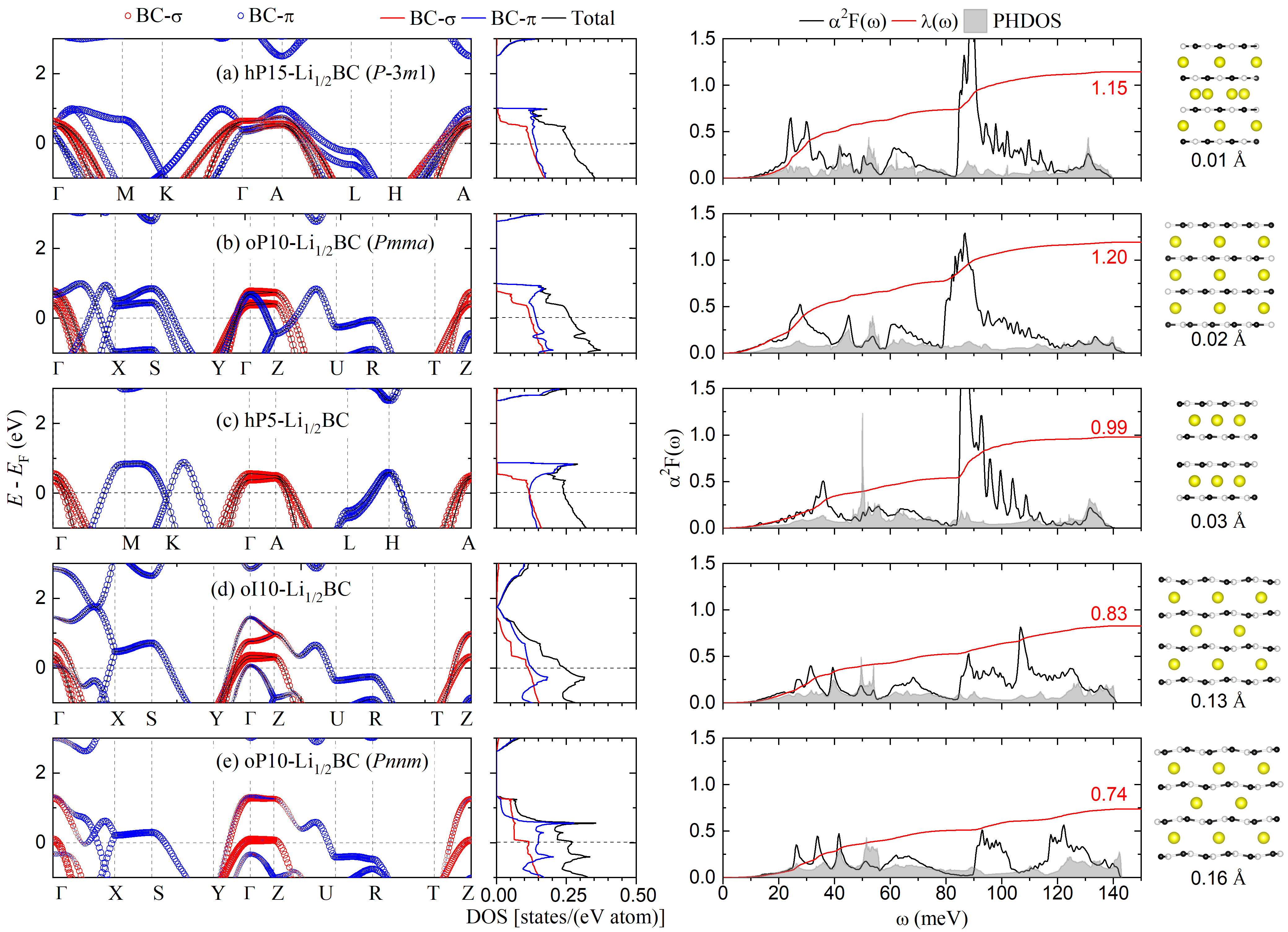}
	\caption{\label{fig-05} Properties of five Li$_{1/2}$BC layered configurations, with the degree of BC layer distortion increasing from (a) to (e). Each row contains the electronic band structure and DOS with orbital characters (two left panels), the phononic DOS, the Eliashberg spectral function $\alpha^2F(\omega)$ with the integrated e-ph coupling strength $\lambda(\omega)$ (right panel), and structural models with the buckling size.}
\end{figure*}

To appreciate the impact of different geometric factors on the material’s superconducting properties, we focused on the single $x=1/2$ composition near the lowest experimentally achievable limit of Li content at which the compound retains the stoichiometric honeycomb BC layers. The most natural deviation from the ideal planar morphology, and apparently the only one supported by measurements, is the corrugation of the covalent layers deduced from Raman and PXRD characterizations of LiBC/Li$_x$BC~\cite{Hlinka2003,Kalkan2019}. It has been pointed out that the BC layers are ‘floppy’ and, just like in Mg$_{1/2}$BC~\cite{worle1994}, tend to pucker around metal vacancies and/or stacking faults~\cite{Hlinka2003,Kalkan2019}. The observed behavior is consistent with previous~\cite{ Kalkan2019,Kharabadze2023} and current DFT results showing that structures with uneven BC layers are thermodynamically favorable. In addition to the most stable oP10$^{\rm SRO}$ ($Pnnm$) and oI10$^{\rm SRO}$ configurations with high degree of buckling, we considered hP5$^{\rm RH}$, oP10$^{\rm SRO}$ ($Pmma$), and hP15$^{\rm SRO}$ ($P\bar{3}m1$) models with low out-of-plane distortions (see Fig.~\ref{fig-05} and Table~S1~\cite{SM}). We defined the buckling magnitude as the deviation from the mean for the B and C atom vertical positions in each layer and found it to range from 0.01~\AA\ in hP15$^{\rm SRO}$ ($P\bar{3}m1$) to 0.16~\AA\ in oP10$^{\rm SRO}$ ($Pnnm$).

The hexagonal-to-orthorhombic symmetry breaking has a pronounced effect on the electronic states shown in Fig.~\ref{fig-05}. In all three orthorhombic models, the important $\sigma$ bands split at the $Z$ point by up to 1~eV, and in oI10$^{\rm SRO}$ the upper $\sigma$ pair additionally splits at $\Gamma$ due to the asymmetric distortion pattern that shortens half of the distances between B-B and C-C atoms in adjacent layers (Fig.~\ref{fig-05}). While the resultant band energy shifts have a moderate impact on the $N_{\sigma}(E_{\textrm{F}})$ contribution because of the quasi-2D nature of the $\sigma$ states, the markedly different hole-doping levels lead to substantial variations in the size of the signature hourglass-shaped Fermi surfaces along the $\Gamma-Z$ line. In particular, the nearly filled bottom pair of the bands in oP10$^{\rm SRO}$ ($Pnnm$) generates Fermi surfaces with dramatically reduced radii (Fig.~S5~\cite{SM}).

The layer distortions also reshape the isotropic Eliashberg spectral function around the frequencies of the bond-stretching B-C modes known to couple strongly with the $\sigma$ states. In the structures with low buckling (Fig.~\ref{fig-05}(a-c)), the profile is dominated by a sharp peak around 90~meV that tapers off at higher frequencies. The large BC layer corrugation causes the peak to split into two broader lower-intensity peaks centered around 90~meV and 110--120~meV (Fig.~\ref{fig-05}(d,e) and Fig.~S6~\cite{SM}). The reduction of the integrated e-ph coupling strength in the two sets of considered structures arises from the weaker coupling in both regions below and above the 90-meV peak: 0.75+0.40=1.15, 0.75+0.45=1.20, and 0.55+0.44=0.99 in hP15$^{\rm SRO}$ ($P\bar{3}m1$), oP10$^{\rm SRO}$ ($Pmma$), and hP5$^{\rm RH}$ with nearly flat layers versus 0.55+0.28=0.83 and 0.55+0.19=0.74 in the significantly buckled oI10$^{\rm SRO}$ and oP10$^{\rm SRO}$ ($Pnnm$) configurations.

By solving the aME equations~\cite{Margine2013}, we obtained the energy distribution of the superconducting gap $\Delta_\bk$ as a function of temperature to find the $T_{\rm c}$ of each structure, shown in Fig.~S7~\cite{SM}. We observed a substantial near-linear decrease from $\sim$ 71~K to $\sim$ 32~K (at $\mu^*=$0.20) in the $T_{\rm c}$ values going from the least to most buckled configurations. The anisotropic domes in the structures with essentially flat BC honeycomb layers are well separated, which is consistent with the superconductivity two-gap profiles found in the MgB$_2$ and LiB materials with planar B frameworks~\cite{Kafle2022}. In contrast, the lower energy gap in the corrugated structures has a considerably larger spread and almost reaches the region of the higher energy gap peaks. The findings indicate that BC layer distortions are indeed detrimental to superconductivity but not sufficient to suppress it entirely, as summarized in Fig.~\ref{fig-06}(b).

One can anticipate the bond rotation and interlayer bridging transformations, favored around $x=1/2$, to have more drastic impacts on the material’s electronic and vibrational properties. We chose four representative lowest energy structures to examine the two defect types: oP20-Li$_{1/2}$BC$^{\rm IB}$ with AA-stacking, mP20-Li$_{1/2}$BC$^{\rm IB}$ with AA’-stacking, mP15-Li$_{1/2}$BC$^{\rm BR}$, and mP30-Li$_{1/2}$BC$^{\rm BR}$. In the bridged configurations, one can still identify hole-doped $\sigma$ states but their contribution to the DOS at the Fermi level is reduced significantly (Fig.~S8(a,b) and Table S1~\cite{SM}). The Eliashberg spectral function shapes have certain similarities with those found for the most distorted SRO structures at this composition, with the major peak around 90 meV either split or shifted up to about 120~meV, but the overall $\lambda$ values are reduced by more than a factor of two down to 0.2--0.3 (Fig.~S9(a,b)~\cite{SM}). Given the small e-ph coupling strengths, we did not attempt aME-level calculations of the $T_{\rm c}$ and found the iME values to be under 1~K. 

In mP15-Li$_{1/2}$BC$^{\rm BR}$, the bond rotation causes the $\sigma$ states to drop below the Fermi level by almost 1~eV leaving only the $\pi$-type Fermi surfaces (Fig. S8(c)). The observation is consistent with the insightful explanation of the electronic DOS response to B-C swap offered by Fogg et al.~\cite{Fogg2006}. We further investigated the bond rotation impact on the electronic structure by constructing a larger mP30$^{\rm BR}$ unit cell with diluted concentration of the defect. The top of the $\sigma$ manifold remained at $-1$~eV but the structure developed a $\sigma$ band above the Fermi level slightly dispersed between 0.5~eV and 1~eV (Fig.~S8(d)~\cite{SM}). Notably, these structures with direct C-C bonds have harder phonon modes reaching 170~meV, which may be used to detect the defects. With no $\sigma$ Fermi surfaces present in either mP15$^{\rm BR}$ or mP30$^{\rm BR}$, the materials have even lower $\lambda \sim 0.2$, respectively, and negligible $T_{\rm c}$ values (Fig.~S9(c,d)~\cite{SM}). We conclude that the appearance of such defects in delithiated structures can indeed explain the reported lack of detectable superconductivity in these intriguing ternary compounds.

\subsection{Stability of \texorpdfstring{Li$_x$M$_y$BC}{Lg}}

\begin{figure*}[t!]
   \centering
\includegraphics[width=0.99\textwidth]{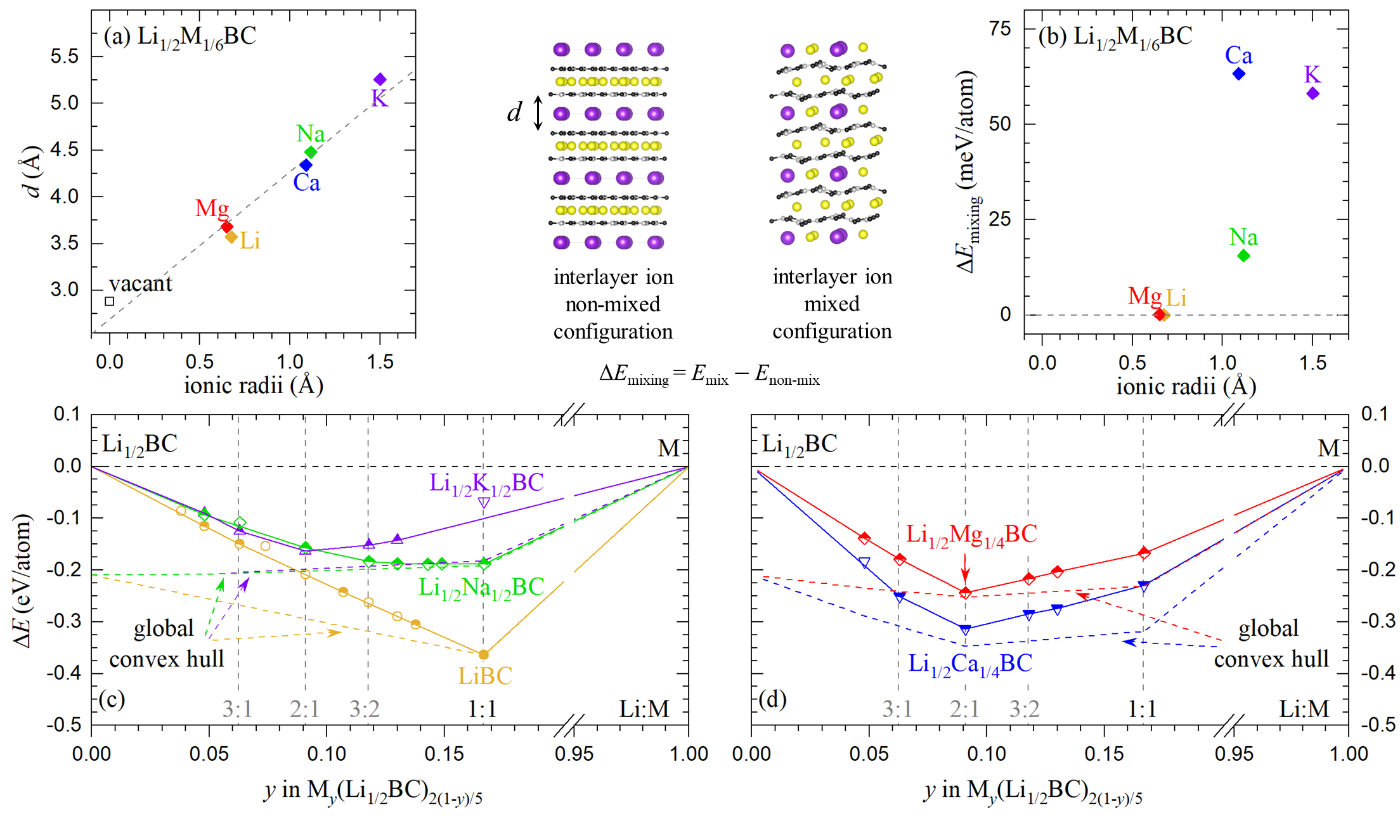}
\vspace{-4mm}
    \caption{\label{fig-07}  Structural and stability properties of double-metal borocarbides. (a) Dependence of the interlayer distance on the intercalant ion size~\cite{radii} at the 3:1 Li:M ratio. (b) Energy difference between configurations with mixed and segregated distributions of extra metals. (c,d) Relative energies of Li$_{1/2}$M$_{5y/[2(1-y)]}$BC phases referenced to oI10-Li$_{1/2}$BC and pure M. The stable LiBC is marked with a solid circle. Metastable phases defining the local convex hull in the kinetically-protected subspace of layered borocarbides are shown as half-filled points connected with solid lines. The global convex hulls are displayed with dashed lines.}
\end{figure*}

The rationale for adding another metal into the metastable delithiated phases is to help recover the ordered layered morphology of the starting LiBC material by suppressing or healing possible defects in the BC honeycomb layers while keeping the bonds hole-doped. Relying on the fact that the covalent framework in Li$_x$BC withstands temperatures exceeding 1,500$^\circ$C, we constrained the search for viable Li$_x$M$_y$BC phases to the honeycomb layered structures that may be kinetically accessible through reintercalation. We focused on the Na, K, Mg, and Ca alkali/alkaline earth metals, closest to Li in terms of size and valence, to investigate whether the insertion of the metals is a thermodynamically downhill reaction. To efficiently explore the considerably expanded configuration space with $x\ge y$ and $1\ge x+y\ge 1/2$, we first generated all possible unique decorations of metal sites for 6 unit cells with up to 48 atoms. Our examination of metal distribution stability revealed that only Mg could co-exist with Li in the same galleries, with the larger Na, K, and Ca showing a clear trend to segregate into M-only and Li-only intercalant layers (Fig.~\ref{fig-07}(b)). Based on this observation, we were able to probe much larger unit cells with up to 72 atoms for specific compositions, such as Li$_{1/2}$Na$_y$BC with $1/2>y$.

Figure~\ref{fig-07}(c,d) summarizes our DFT results on the stability of quaternary compounds relative to the starting lowest-energy oI10-Li$_{1/2}$BC$^{\textrm{SRO}}$  and pure M. Naturally, the majority of the considered double-metal borocarbides have the most pronounced stabilization, by $\sim 0.2$--0.3~eV/atom, at compositions that restore the 8-electron count (1:1 for monovalent and 2:1 for divalent metals). Compounds with the largest K are an exception, achieving a comparable relative energy of $-0.16$~eV/atom at a lower total metal content, Li$_{1/2}$K$_{1/4}$BC. Importantly, all four quaternary systems feature phases, denoted with half-filled symbols, that form a convex hull with respect to the endpoints and include both electron-doped (e.g., Li$_{1/2}$Mg$_{1/2}$BC and Li$_{1/2}$Ca$_{1/2}$BC) and hole-doped (e.g., Li$_{1/2}$Na$_{1/4}$BC and Li$_{1/2}$Ca$_{1/6}$BC) derivatives (similar profiles can be seen for other fixed Li concentrations up to $x=3/4$ in Fig.~S10~\cite{SM}).

One can expect the formation of these locally stable phases to depend strongly on the morphology of the starting Li$_{1/2}$BC. If each gallery in the ternary compound is half-filled, as our DFT calculations indicate, then intercalation of the extra metal would result in mixed-metal structures suboptimal for all but Li-Mg-BC quaternaries. However, even these heterogeneous configurations are well below the oI10-Li$_{1/2}$BC-M tie-line (e.g., for Li$_{1/2}$M$_{1/6}$BC in Fig.~\ref{fig-07}(b,c,d)) meaning that there would be a thermodynamic force driving the larger metals into the ternary matrix. The ensuing expansion of the interlayer distances would likely increase the migration of the Li ions~\cite{Kang2006} and could lead to an eventual metal segregation through the sample edges. If Li$_{1/2}$BC does have the starting stage-2 morphology, the larger metals would need to overcome a higher barrier to enter the empty galleries. Fig.~\ref{fig-07}(a) shows that the interlayer spacing correlates well with the ion size and would have to increase considerably, e.g., nearly double from 2.8~\AA\ in hP5-Li$_{1/2}$BC$^{\rm RH}$ to 5.3~\AA\ in Li$_{1/2}$K$_{1/6}$BC, to accommodate the extra alkali or alkaline earth metals.

The potential advantage of such quaternaries over Li$_x$BC can be seen from the following analysis of the defect stability in various compounds. To understand the importance of the size and electron count factors, we first simulated the aforementioned rotated bond and interlayer bridge defects in the {\it ternary} analogs with half-filled metal sites. Fig.~S11(c)~\cite{SM} illustrates that both defective structures become unstable by at least 100~meV/atom in the electronically neutral M$_{1/2}$BC compounds (with M = Mg or Ca) and get significantly disfavored in the hole-doped M$_{1/2}$BC (with M = K). We next evaluated the relative stability of the defective structures with similar contents but different types of the intercalant metals, e.g., Li$_{2/3}$BC versus Li$_{1/2}$M$_{1/6}$BC. The overall trend, particularly prominent for bond rotations, is a destabilization of the non-honeycomb BC morphologies in the quaternary compounds with the larger and more electron-rich metals. It should be kept in mind that the imperfections appearing in the covalent layers during high-$T$ delithiation are likely protected by high kinetic barriers and might not easily heal upon insertion of additional metals but, being less stable in the quaternary compounds, they could be more responsive to annealing.

\begin{figure*}[t!]
   \centering
\includegraphics[width=0.95\textwidth]{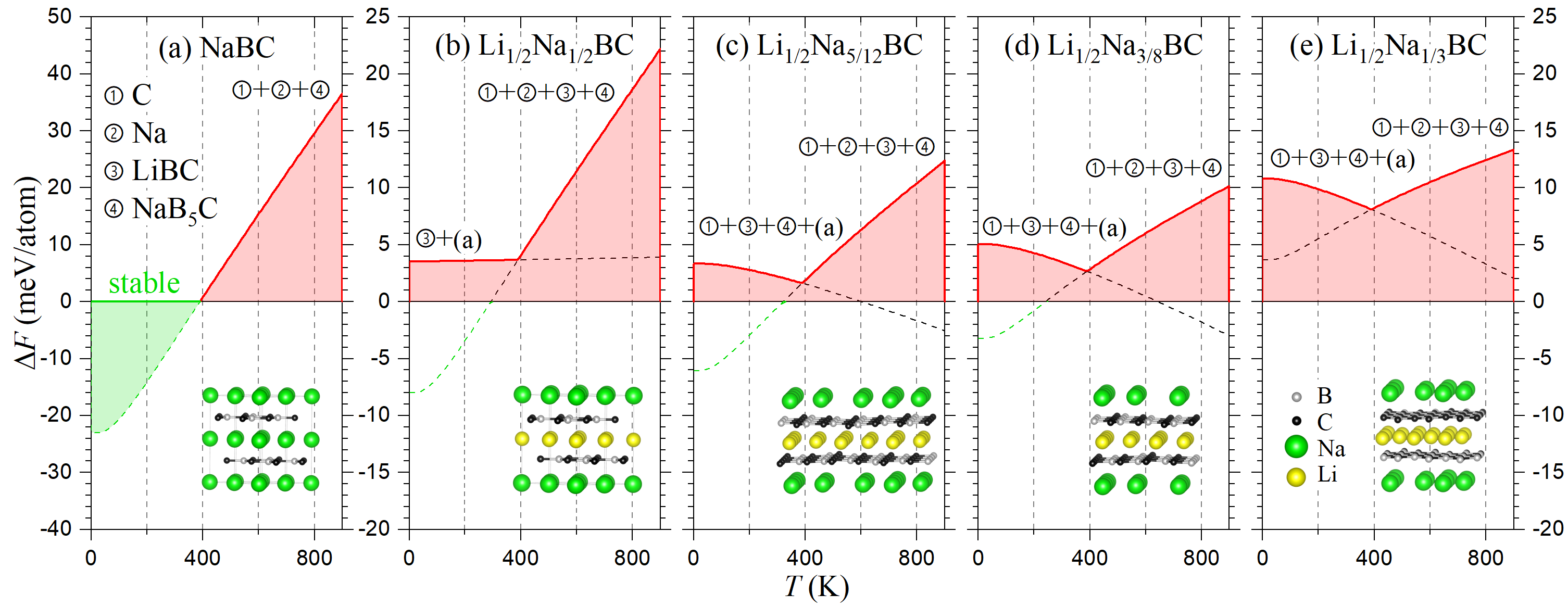}
    \caption{\label{fig-08} Distance to the convex hull as a function of temperature for proposed Na and Li-Na borocarbides. Details on the hP6-Li$_{1/2}$Na$_{1/2}$BC, mP35-Li$_{1/2}$Na$_{5/12}$BC, mP23-Li$_{1/2}$Na$_{3/8}$BC, and mP34-Li$_{1/2}$Na$_{1/3}$BC phases can be found in Table~\ref{table1}. Different lines correspond to the lowest-free energy combinations among all observed ground states (circled numbers) and proposed phases (bracketed letters), with the full set of previously synthesized Li-Na-B-C phases listed in Fig.~S12~\cite{SM}. NaBC and Li$_{1/2}$Na$_y$BC ($y=$ 1/2, 5/12, and 3/8) appear below the {\it known} convex hull at low temperatures, which is illustrated with dashed green lines. In particular, NaBC is thermodynamically stable up to 400~K, being below a facet of the convex hull defined by the observed C, Na, and NaB$_5$C phases, while hP6-Li$_{1/2}$Na$_{1/2}$BC is below the C, Na, LiBC, and NaB$_5$C mixture up to 300~K (the free energy of Na evaluated for the solid state even above the melting point of 371 K represents an upper bound). Although the three quaternary phases in panels (b-d) become metastable upon inclusion of the predicted NaBC in the optB86b calculations, mP23-Li$_{1/2}$Na$_{3/8}$BC is globally stable according to the optB88 results (Fig.~S14~\cite{SM}).}
\end{figure*}

\begin{figure}[b!]
   \centering
\includegraphics[width=0.45\textwidth]{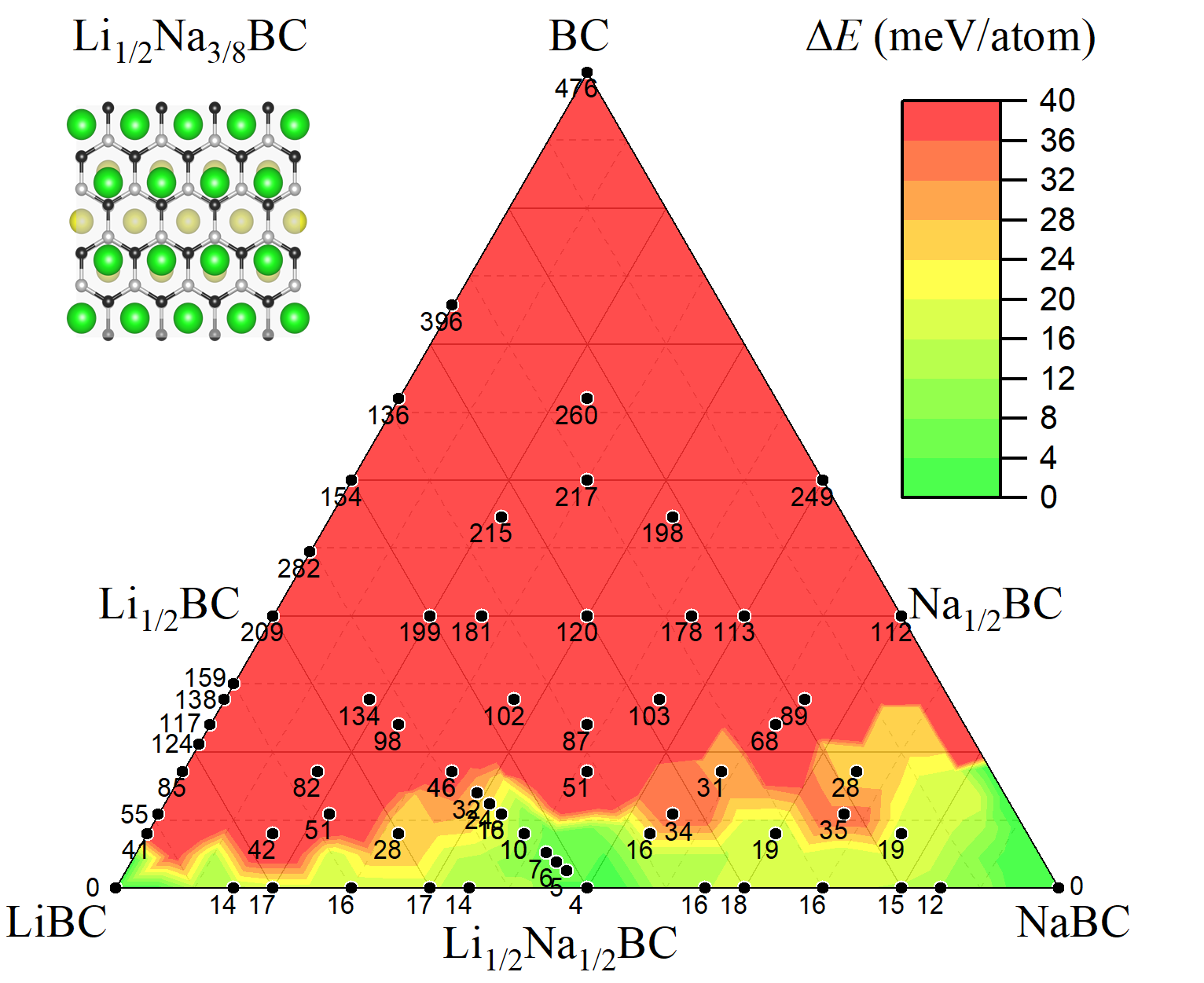}
    \caption{\label{fig-09} Distance to the global convex hull for Li$_x$M$_y$BC phases with honeycomb morphology at $T=0$ K.}
\end{figure}

Finally, we discuss the global stability of the identified quaternary phases by constructing the full convex hulls based on DFT energies of the Li-B-C materials listed in ref.~\cite{Kharabadze2023} and all reported relevant compounds with the four additional metals~\cite{materials-project,Emery2005,Delacroix2021,miao2016}. For phases (near) stable at $T=0$~K, we also calculated free energies by taking into account the vibrational entropy and constructed convex hulls at elevated temperatures. 

In the Li-K-B-C system with the large mismatch of the alkali metal sizes, no quaternary phases with honeycomb BC layers appeared within 35~meV/atom to the boundary of the convex hull (see Figs.~\ref{fig-07}(c) and~S12(b)~\cite{SM}). In the Li-Ca-B-C chemical space, the double-metal layered borocarbides are much more competitive because the large divalent Ca ions do not need to occupy every vacant metal site to make the electron-deficient BC network whole. Unfortunately, this trait has already been observed in the known Ca$_{1/2}$BC compound with a covalent framework comprised of 4-8 polygons~\cite{Albert1999}. The ternary phase makes the quaternary borocarbides metastable by at least 30~meV/atom across the investigated composition range (Figs.~\ref{fig-07}(d) and S12(d)) and could be the end product of the Li$_{1/2}$BC intercalation with Ca if the barrier to rebonding between the two related layered morphologies is not sufficiently high. The Li-Mg-BC metal borocarbides are technically even closer to stability, with some phases being only 8~meV/atom above the boundary of the convex hull (Fig.~\ref{fig-07}(d) and S12(c)). Regrettably, the existence of the Mg$_{1/2}$BC phase with the honeycomb BC layers~\cite{worle1994} means that the phase separation of Li-Mg borocarbides into the corresponding ternaries would be achieved by a simple diffusion of metal ions within galleries, which greatly reduces the chances of obtaining stable ordered quaternary compounds. We are aware of only a few attempts to synthesize mixed-metal borocarbides. Bharathi et al.~\cite{Bharathi2002} observed the formation of an electron-doped Li$_{0.8}$Mg$_{0.2}$BC, while Renker et al.~\cite{renker2004} obtained Li$_{0.95}$Na$_{0.05}$BC. Mori et al.~\cite{Mori2004} reported the successful synthesis of Mg$_{0.25}$Li$_{0.40}$BC, although no superconductivity was observed in their samples down to 1.8~K. Their intended doping of Mg$_{1/2}$BC with B or Li (under high pressure) resulted only in large amounts of neighboring phases MgB$_2$ and LiBC.

The Li-Na-B-C compositional space, on the other hand, may host true ground states that have not yet been observed. The structure search by Miao et al. demonstrated that hP6 is the lowest-energy configuration at the NaBC composition but its global stability in the Na-B-C ternary system was not evaluated~\cite{miao2016}. The Materials Project~\cite{materials-project, mp-1, mp-2, mp-3} presently places this phase (mp-1238775) 38~meV/atom above the convex hull facet defined by C, Na, and B$_4$C but the estimate is unreliable for several reasons. First, the graphite entry (mp-2516584) contains only the Perdew-Burke-Ernzerhof (PBE)~\cite{PBE} total energy and a manual evaluation of the distance based on the reported values in this approximation produces 27.9~meV/atom, close to 25.0~meV/atom in our PBE calculations. Second, none of the presently used Materials Project functionals accounts for the dispersive interactions particularly important in the considered layered materials~\cite{ak06,Lebegue2010,ak30,Ning2022}. Finally, the database is missing the known NaB$_5$C phase experimentally determined to have a simple cP7 structure~\cite{Delacroix2021} but difficult to simulate because of a random distribution of C on the B sites. Having sampled $\sim 200$ unique decorations of different unit cells with up to 35 atoms, we found an oP14 model to have the lowest energy, 23-40~meV/atom below the randomized $P1$ models generated by Delacroix et al.~\cite{Delacroix2021} for calculating NMR spectra. The phase proved to be important in the construction of the Na-B-C convex hull, as the $\frac{4}{15}$C+$\frac{4}{15}$Na+$\frac{7}{15}$NaB$_5$C mixture is $-16$ meV/atom below $\frac{1}{4}$C+$\frac{1}{3}$Na+$\frac{5}{12}$B$_4$C in our optB86b calculations at $T=0$ K. Yet, NaBC is below either reference set, by $-20$ and $-36$~meV/atom, respectively. Considering the diversity of the bonding types and structural motifs defining the convex hull, one can expect the results to be sensitive to the DFT approximation. With the optB88~\cite{optB88} (r2SCAN+rVV10~\cite{Furness2020,Ning2022}) functional, we find that NaBC is located even deeper, at $-49$ and $-74$ meV/atom ($-59$ and $-89$~meV/atom), below the two respective facets. Fig.~\ref{fig-08}(a) shows that the vibrational entropy quickly destabilizes this ternary phase, which may explain why it has not been produced so far. 

The double-metal Li$_{1/2}$Na$_{1/2}$BC compound in the simple hP6 structure also appears stable relative to C, Na, NaB$_5$C, and LiBC at $T=0$ K but is less favorable with respect to LiBC and NaBC at all temperatures (Figs.~\ref{fig-08}(b)-\ref{fig-09}). It is evident that for this midpoint quasi-ternary phase the Li and Na size mismatch induces an elastic energy penalty. We investigated the interplay between the strain and the electron count by considering Na-deficient compounds shown in Fig.~\ref{fig-08}(c-e). The nearly fully-filled Li$_{1/2}$Na$_{5/12}$BC and Li$_{1/2}$Na$_{3/8}$BC (see Figs.~\ref{fig-08} and~\ref{fig-09}) prefer to have missing Na rows along the zigzag direction, while mP34-Li$_{1/2}$Na$_{1/3}$BC achieves the highest stability via stretching the Na sublattice to skip one hexagon row along the armchair direction. Fig.~\ref{fig-08}(c) illustrates a clear benefit of removing a small amount of the larger metal, which leads to not only a slightly lower energy at $T=0$ K in Li$_{1/2}$Na$_{5/12}$BC but also a lower free energy at elevated temperatures due to the softening of Na phonon modes in the less-crowded hole-doped material (Fig.~S13~\cite{SM}). Therefore, there is likely a range of temperatures around a few hundred K and $1/2>y$ where Li$_{1/2}$Na$_y$BC become true ground states, as our optB88 calculations indicate for $y=3/8$ (Fig.~S14~\cite{SM}), and may form from the elements. 

It should be noted that in the important NaB$_5$C compound the configuration entropy due to the disordered occupation of the B sites by C can noticeably lower the {\it free energy} at high temperatures. However, our comparison of the adopted oP14 representation with a considerable deviation from the cubic shape ($b/a=0.936$) against a far more uniform but low-symmetry aP189 model~\cite{Delacroix2021} revealed a significant {\it energy} penalty of $\sim 30$~meV/atom for NaB$_5$C to have the randomized B$_5$C covalent framework that apparently remains frozen at low temperatures. In contrast, the proposed Na-deficient Li$_{1/2}$Na$_y$BC phases may also benefit from the configurational entropy due to Na disorder at high temperatures but should be able to become ordered through ion migration upon annealing. It is also important to acknowledge that identification of materials thermodynamically stable with respect to known ones only indicates that {\it additional} phases, not necessarily the predicted ones, should exist in the considered chemical system. Indeed, no search algorithm can guarantee the identification of the full set of true ground states, especially in vast multielement configuration spaces.

\subsection{Superconductivity of \texorpdfstring{Li$_x$M$_y$BC}{Lg}}

In the final set of calculations, we examined whether the quaternary layered borocarbides identified as viable (meta)stable Li$_x$M$_y$BC materials retain the desired electronic and vibrational features to be good superconductors. The phases chosen for this analysis are the lowest-energy stage-2 hP5-Li$_{1/2}$BC$^{\rm RH}$ derivatives with extra metals reintercalated into the expanded empty galleries at different Li:M ratios (4:3, 3:2, 2:1, or 3:1 for group-I and 2:1 or 3:1 for group-II metals). The structural and electronic information of the six hole-doped ($n$ = $+1/3$, $+1/4$, $+1/6$, and $+1/8$ holes/u.c.) and two electron-doped ($n$ = $-1/6$ and $-1/2$ electrons/u.c.) phases is given in Table~S1 and Figs.~S17 and S19~\cite{SM}. The electron doping levels achieved in Li$_{1/2}$Ca$_y$BC for $y=$ 1/3 and 1/2 leads to the population of non-coupling interstitial states and generates 3D Fermi surfaces (see Fig.~S17~\cite{SM}). Unsurprisingly, the resulting $T_{\rm c}$ of $\sim 2$~K obtained in the iME calculations are much lower than those predicted for the corresponding hole-doped Li$_x$BC candidates.

The band structure and DOS plots in Fig.~S19~\cite{SM} indicate that the hole-doped quaternary phases retain the signature 2D $\sigma$ bands. Keeping the electron count constant by considering hP17-Li$_{1/2}$Na$_{1/3}$BC, Li$_{1/2}$Mg$_{1/6}$BC, and Li$_{1/2}$Ca$_{1/6}$BC compositions against the control case of Li$_{5/6}$BC, we could probe the response of the electronic structure to the presence of different metals. The $\sigma$-state contributions to $N(E_{\textrm{F}})$ remain the same in the Na and Ca quaternaries and increase slightly in the Mg counterpart, while the $\pi$-state contributions show a larger variability of $\sim 30$\% because of the different interlayer expansions and metal ion contents. The scan across the Li$_x$M$_y$BC series with doping levels ranging from $+1/8$ to $+1/3$ holes/u.c. reveals $N(E_{\textrm{F}})$ variation by $\sim 40$\%, from 0.20~states/(eV atom) in hP17-Li$_{1/2}$Na$_{1/3}$BC to 0.29~states/(eV atom) in Li$_{1/2}$K$_{1/6}$BC, with no clear trend. 

\begin{figure}[t]
	\centering
    \includegraphics[width=0.49\textwidth]{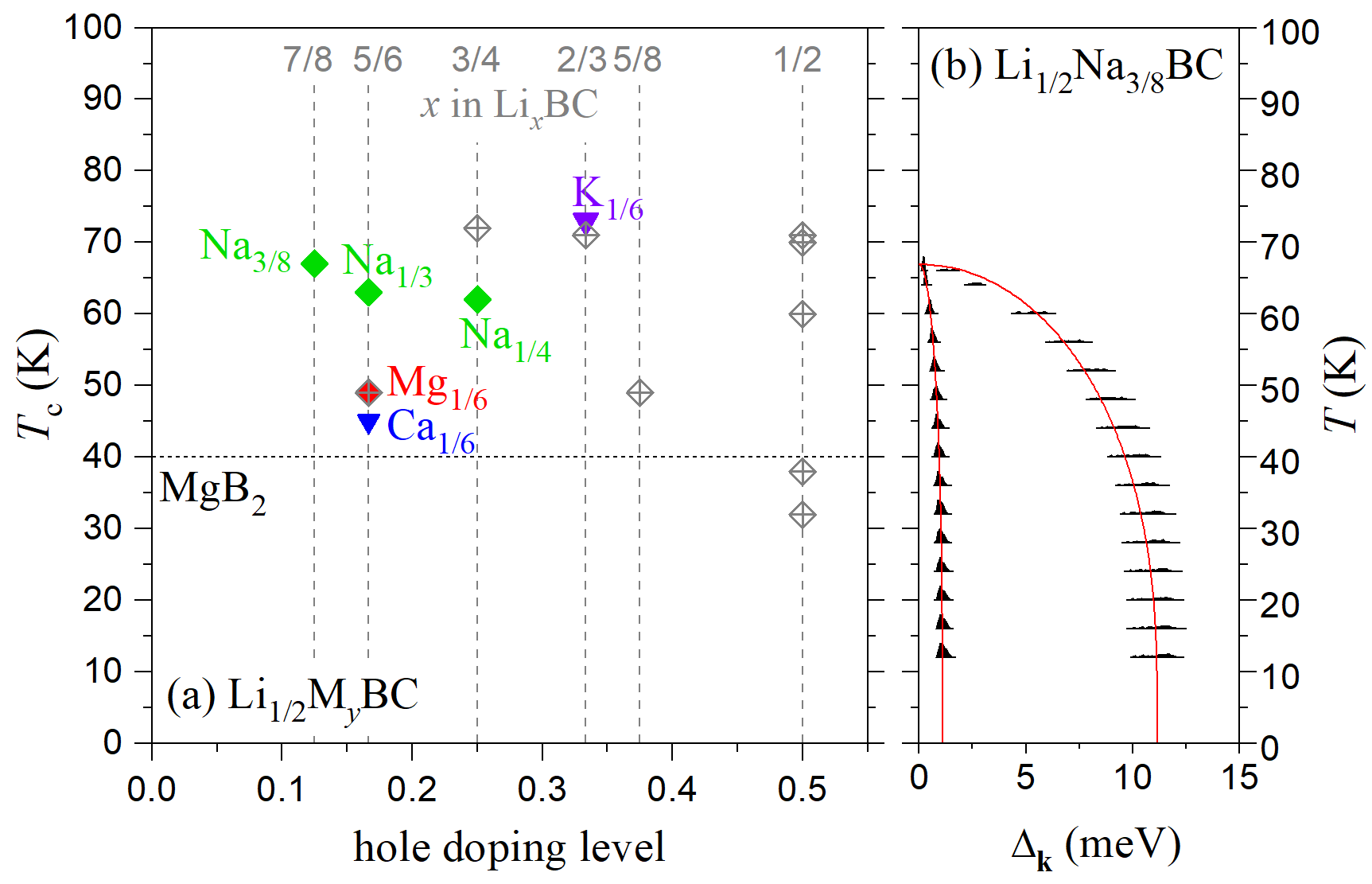}
	\caption{\label{fig-10} Superconducting properties of metal borocarbides calculated at the aME level with $\mu^*=0.20$. (a) $T_{\rm c}$ for select Li$_x$BC (grey diamonds) and Li$_{1/2}$M$_y$BC (other symbols) phases as a function of hole doping. For comparison, the $T_{\rm c}$ value of 40 K found for MgB$_2$ at this level of theory is displayed with a dotted line. (b) Superconducting gap in Li$_{1/2}$Na$_{3/8}$BC.}
\end{figure}

The similarities between the ternary and quaternary hole-doped borocarbides extend to the calculated Eliashberg spectral functions shown in Fig.~S20~\cite{SM}. Since the Li phonon modes couple with electrons weakly based on our $\lambda$-resolved analysis, the frequency drop of the heavier metal vibrations has little effect on the total e-ph coupling strength. The dominant peaks present in all materials correspond to the in-plane bond-stretching B-C modes, with the highest intensity peak occurring in the same range of frequencies as seen in the Li$_x$BC ternary. The aME results summarized in Fig.~S21~\cite{SM} demonstrate that the quaternary phases preserve the two highly anisotropic superconducting gaps. The predicted $T_{\rm c}$ range from 45~K in Li$_{1/2}$Ca$_{1/6}$BC to 73~K in Li$_{1/2}$K$_{1/6}$BC (at $\mu^*=0.20$), comparable to the highest values obtained for the ternary Li$_x$BC phases (see Fig.~\ref{fig-10}).

Compared to MgB$_2$, the quaternary borocarbides have similar 0.23 states/(eV atom) values of $N(E_{\rm F})$, as shown in Figs.~S19, S20, and S23~\cite{SM}. The overall higher e-ph coupling strength in the latter can be attributed to the contribution of low-frequency BC modes between 40-50 meV. For instance, below the largest peak in $\alpha^2F(\omega)$, the modes contribute an average 42\% to total $\lambda$ in Li$_{1/2}$M$_y$BC, while only about 24\% in MgB$_2$. 

\section{Summary}
\label{sec:conclusions}

The presented first-principles reexamination of Li$_x$BC provides new insights into the seemingly well-studied material once considered to be one of the most promising high-temperature ambient-pressure conventional superconductors. Despite the extensive synthesis and characterization work~\cite{Bharathi2002, Fogg2003a, Fogg2003b, Zhao2003, Fogg2006, Kalkan2019}, our analysis indicates that the precise structures of some Li$_x$BC derivatives obtained via delithiation at high temperatures remain unresolved. To identify possible forms of the metastable hole-doped phases, we employed a combination of structure search and analysis strategies (Fig.~\ref{fig-02}). Our modeling of domain configurations has indicated that the stage-2 model proposed to explain PXRD data~\cite{Kalkan2019} is not only energetically disfavored but also inconsistent with the expected interlayer spacing dependence on the Li content (Fig.~\ref{fig-04}). The evolutionary searches have uncovered significantly more stable non-planar BC morphologies at $x=1/2$ including configurations with fully-connected 3D frameworks. The combinatorial screening of possible Li compositions and decorations has helped establish that the alternative BC motifs become favored sooner than previously suspected~\cite{Fogg2006,Kharabadze2023}, around $x=5/8$ (Fig.~\ref{fig-03}). Even though the reported PXRD patterns clearly indicate that Li$_x$BC retains its hexagonal layered structure down to $x=1/2$, the rebonded patterns may appear in the form of local defects. We have used aME calculations to quantify the effect of structural transformations on the critical temperature. Surprisingly, the lowest-energy ordered layered configurations have been found to have high $T_{\rm c}$ across the full examined $5/6\ge x\ge 1/2$ range. The natural layer buckling has a relatively moderate effect on the superconducting properties, reducing the highest $T_{\rm c}$ of $\sim$ 70~K by roughly a factor of two (Table~\ref{table1}). The interlayer bridging suppresses $T_{\rm c}$ down to a few Kelvin, while the intralayer B-C bond rotation extinguishes the phonon-mediated superconductivity completely (Fig.~\ref{fig-06}). The absence of superconductivity in annealed samples~\cite{Zhao2003} suggests that such defects cannot be completely eliminated, a factor that diminishes the prospect of high-$T_{\rm c}$ superconductivity in the delithiated ternary borocarbide.

A possible solution examined in this study is to repopulate vacant metal sites with other metals via soft reactions constrained by the layered honeycomb morphology. We have determined that the defects detrimental for superconductivity have higher energies of formation if galleries are filled with larger ions and, hence, the double-metal borocarbides might be more amenable to annealing than the starting Li$_x$BC material. We have shown that the intake of Na, K, Mg, or Ca is thermodynamically favorable along the kinetics-restricted pathway and could lead to formation of various hole-doped phases, such as Li$_{1/2}$K$_{1/6}$BC or Li$_{1/2}$Ca$_{1/6}$BC (Fig.~\ref{fig-07}), with estimated $T_{\rm c}$ up to 73~K (Fig.~\ref{fig-10}). Borocarbides based on Na appear to be the most promising. Our DFT calculations indicate that Li$_{1/2}$Na$_y$BC with $y$ just below 1/2 could be true ground states and, hence, synthesizable from the elements (Fig.~\ref{fig-08}). In contrast to Li$_x$BC, these phases are naturally hole-doped and could be long-sought-after ambient-pressure materials with high-$T_{\rm c}$ conventional superconductivity. We have also demonstrated an unexpected thermodynamic stability of the ternary NaBC analog of LiBC at low temperatures. If the semiconducting compound does form, it would be interesting to investigate whether it could be desodiated and turned into a high-$T_{\rm c}$ superconductor. The findings indicate that layered metal borocarbides comprise a materials class uniquely suited to host viable ambient-pressure conventional superconductors rivaling the MgB$_2$ archetype.

\section*{Author Contributions}
C. R. Tomassetti and G. P. Kafle contributed equally to this work. 

\section*{Conflicts of interest}
There are no conflicts to declare.

\section*{Acknowledgements}

The authors acknowledge support from the National Science Foundation (NSF) (Awards No. DMR-2320073 and DMR-2132586). This study used the Frontera supercomputer at the Texas Advanced Computing Center through the Leadership Resource Allocation (LRAC) award DMR22004. Frontera is made possible by NSF award OAC-1818253~\cite{Frontera}. This work also used the Expanse system at the San Diego Supercomputer Center through allocation TG-DMR180071 from the Advanced Cyberinfrastructure Coordination Ecosystem: Services \& Support (ACCESS) program~\cite{ACCESS}, which is supported by NSF grants \#2138259, \#2138286, \#2138307, \#2137603, and \#2138296. The authors thank Igor Mazin for helpful discussions.


\bibliography{refs} 
\bibliographystyle{rsc}

\end{document}